\documentclass[journal]{IEEEtran}

\usepackage{wrapfig,enumerate,color,url}
\usepackage{tabularx}

 \usepackage{graphicx}
  \graphicspath{{./Figures/}}
\usepackage[cmex10]{amsmath}
\usepackage{amssymb,comment}

\usepackage{epstopdf}

\usepackage[tight,footnotesize]{subfigure}

\newcommand{\figurewidth}{\columnwidth}

\begin{document}

\title{State-of-the-art in Power Line Communications: from the Applications to the Medium}

\author{Cristina Cano$^1$, Alberto Pittolo$^2$, David Malone$^3$, Lutz Lampe$^4$ Andrea M. Tonello$^5$, Anand Dabak$^6$,\\
\small
$^1$Trinity College Dublin, Ireland,
$^2$University of Udine, Italy,
$^3$Maynooth University, Kildare, Ireland,
\\
$^4$University of British Columbia, Vancouver, Canada,
$^5$University of Klagenfurt, Austria,
$^6$Texas Instruments Inc., Dallas, USA.
}

\maketitle

\begin{abstract}

In recent decades, power line communication has attracted considerable attention from the research community and industry, as well as from regulatory and standardization bodies. In this article we provide an overview of both narrowband and broadband systems, covering potential applications, regulatory and standardization efforts and recent research advancements in channel characterization, physical layer performance, medium access and higher layer specifications and evaluations. We also identify areas of current and further study that will enable the continued success of power line communication technology.

\end{abstract}


\IEEEpeerreviewmaketitle

\section{Introduction}\label{sec:intro}

The use of electrical wires to provide data transmission capabilities, known as Power Line Communication (PLC), has recently experienced increased deployment. Chip manufacturers of PLC devices for in-home and for smart grid applications report that they are shipping millions of such devices each year and expect the number to continue to grow in the future. 

PLC networks provide a number of advantages that make them both a useful complement and a strong competitor to wireless networking solutions. The main appeal of PLC networks is their low deployment cost when an electrical wired infrastructure is already in place. In addition, PLC networks allow communication through obstacles that commonly degrade wireless signals, while delivering high data-rates. Moreover, PLC also provides a low-cost alternative to complement existing technologies when aiming for ubiquitous coverage. For instance, as a backhaul for wireless sensor networks or small cells.

Since one of the main advantages of using PLC networks is the possibility of re-using the existing wired electrical network to provide communication capabilities, the smart grid remains one of the most appealing applications of PLC and consequently the research carried out in this area is vast. Some feasibility and experimental studies include the works in \cite{pinomaahomeplug,papilaya2014analysis,pinomaa2014applicability,goedhart2012adapting,pinomaa2011power,he2010ict,liu2010application}. In the same line, smart city \cite{mlynek2013measurements}, in-home automation \cite{dickmann2011digitalstrom} and telemetry \cite{castor2014experimental} applications can naturally benefit from the fact that new cabling is not required and that wireless propagation issues are avoided. 

The high data-rates that can currently be achieved with PLC --- comparable with WiFi and domestic Ethernet --- make it suitable for in-home multimedia applications. These scenarios, along with the smart grid cases, correspond to one of the most studied areas of applicability of PLC networks \cite{wu2012experimental,gnazzo2011powerline,yeh2008available,lin2006robust}.

Applications that provide a means for communication in transport systems where an electrical deployment is already in place can also take advantage of PLC networking \cite{degauque2015power}, \cite[Ch.~10]{lampe2015power}. In this context, PLC networks have been explored for use in in-vehicle communications \cite{tanguy2009power,barmada2010power}, naval \cite{antoniali2011measurements} and aircraft systems \cite{degardin2010possibility}, as well as in trains \cite{barmada2008design}.

However, the applicability of PLC networks is not restrained to these scenarios. A range of novel applications have been proposed for PLC networks including robotics \cite{minamiyama2011power}, authentication \cite{sherman2010location}, security systems in mining \cite{zhang2014novel}, as well as uses within inductive coupling \cite{tsuzuki2015feasibility}, contactless communication \cite{debeer2014contactless} and wireless power transfer \cite{barmada2014power}.

Given the wide range of applications for which PLC networks can prove useful and the number of associated challenges, PLC has gathered substantial attention from the research community as well as industry and has fostered a range of regulatory and standardization efforts. In this article, we provide a comprehensive overview of the regulation and standardization processes, we summarize the different research questions that have been studied (from the physical layer to higher layers in the stack) and we outline important future research directions, for both narrowband (NB) and broadband (BB) systems.

The remainder of this article is organized as follows. In Section~\ref{sec:standards} we give an overview of the regulatory and standardization activities, showing the history of how current standards arose and their implementation. In Section~\ref{sec:plc_medium} we describe the pertinent characteristics of the PLC channel, including single-input-single-output (SISO) and multiple-input-multiple-output (MIMO) channels, modeling of the channel response, line impedance and noise properties. In Section~\ref{sec:phy} we summarize the research effort assessing the performance of the physical layer, outlining main results and describing potential improvements. In Section~\ref{sec:mac} we describe the MAC protocols defined in the standards and suggest future research directions and areas that are not deeply explored at present. We conclude the article with some final remarks in Section~\ref{sec:remarks}.

\section{Regulation, Standardization Activities and Industrial Solutions}\label{sec:standards} % Lutz

The diversity of grid and application domains to which PLC can be applied has naturally led to a large ecosystem of specifications, many of which have been adopted by standards-developing organizations (SDOs). Regulatory activities are essentially concerned with coexistence with other systems that also use the power grid (i.e. machines and appliances that draw electricity) and wireless systems operating in the same frequency bands as PLC. The frequency range used for today's PLC solutions starts as low as 125~Hz and reaches as high as 100~MHz. A useful classification of PLC systems according to frequency bands has been introduced in \cite{LL-galli:2011}: it distinguishes between ultra-narrowband (UNB), narrowband (NB) and broadband (BB) PLC systems, operating between about 125--3000~Hz, 3--500~kHz and 1.8--100~MHz, respectively. Most recent developments in standardization and regulation activities over the past 20 or so years apply to NB and BB PLC systems, and we will focus on these in the following. 

\subsection{Regulation Activities}

As for any electric load that is connected to the power grid, PLC systems are subject to regulations that limit the strength of the signals coupled into power lines. In most cases, it is desirable that the PLC signal is fully contained within the proximity of the power line. However, since the power grid has not been designed to conduct relatively high-frequency signals, electromagnetic radiation occurs (e.g. \cite{vick2001radiated}). This is mostly relevant for BB PLC systems whose signals  often have short wavelengths compared to the length of the power lines. Hence, the relevant regulatory constraints are generally different for NB and BB PLC systems.

\subsubsection{NB PLC}
\label{sec:nbplcreg}

Europe is a very active market for PLC equipment. The regulation and market access principles for PLC devices as telecommunication equipment are discussed in \cite{LL-Bookcollection:PLC-Ferreira-2010Chap7}, \cite[Ch.~3]{lampe2015power}. Under this framework, European harmonized standards are an accepted approach for product compliance test. An important such standard is the European Norm (EN)~50065, a complete version of which was first issued by CENELEC in 1992 \cite{LL-EN50065-1}. The EN distinguishes four frequency bands, which are commonly referred to as CENELEC-A (3--95~kHz), CENELEC-B (95--125~kHz), CENELEC-C (125--140~kHz) and CENELEC-D (140--148.5~kHz) respectively. It specifies in-band and out-of-band emission limits in terms of maximum voltage levels together with the measurement procedures. Besides band-specific limits, the standard mandates that the CENELEC-A band is reserved for power utilities and that the CENELEC~B-D bands can only be used by consumer installations. Moreover, it 
specifies a mandatory carrier-sense multiple-access with 
collision avoidance (CSMA/CA) mechanism for the CENELEC-C band. EN~50065 has been decisive for the proliferation of NB PLC systems for home and industry automation and for utility use such as smart metering. There is no harmonized standard for frequencies between 150~kHz and 500~kHz yet. However, the IEEE 1901.2 standard specifies conducted disturbance limits in terms of maximal power spectral densities \cite[Sec.~7.5]{LL-IEEE1901.2} and the methods of measurement in an informative appendix \cite[App. E]{LL-IEEE1901.2}, respectively, see also \cite{7147613}.

In the U.S., PLC emissions are regulated through the Code of Federal Regulations, Title 47, Part 15 (47 CFR \S15) by the U.S.\ Federal Communications Commission (FCC) \cite{LL-FCC08}. Here, the regulations for so-called ``power line carrier'' systems in \S15.113 are relevant. This paragraph permits power utilities to use PLC in the  9--490~kHz band ``on an unprotected, non-interference basis''. There is one caveat though, in that these specifications do not apply to ``electric lines which connect the distribution substation to the customer or house wiring''  (\S15.113(f)). Hence, PLC for many smart grid applications involving, for example, smart meters would not fall under this provision. For such cases, one has to consider limits for what is defined as ``carrier current system'' in \S15.3(f). Accordingly, paragraph \S15.107(c) declares that there are only out-of-band conducted emission limits, to protect the 535--1705~kHz band. Furthermore, via paragraphs \S15.109(e) and \S15.209(a), in-band radiated 
emission limits are specified for the frequency range from 9~kHz to 490~kHz. 

Another regulatory document that has been considered in system specifications is the Standard~T84 \cite{LL-ARIB02} by the Japanese Association of Radio Industries and Businesses (ARIB). This permits the use of PLC in the 10--450~kHz band. Table~\ref{tab:nbplcreg} summarizes the emission standards and relevant regulations from the discussed three regions for NB PLC. 

\begin{table}
\begin{center}
\caption{Emission standards and regulations for NB PLC ($<500$~kHz) and BB PLC in different regions of the world.}
\label{tab:nbplcreg}
%\begin{tabular*}{\linewidth}{@{\extracolsep{\fill}}|l||l|l|l|}
\begin{tabularx}{\columnwidth}{|c|p{19mm}|X|}
\hline
Region & \multicolumn{1}{c|}{Standard/}& \multicolumn{1}{c|}{Remarks}\\
& \multicolumn{1}{c|}{Regulation}&\\
\hline\hline
\multicolumn{3}{|c|}{Narrowband PLC}\\\hline\hline
Europe & EN~50065 &  CENELEC~A band for utility use \\
& 3--148.5~kHz& CENELEC~B-D bands for consumer use\\
&& CSMA/CA in CENELEC~C band\\ \cline{2-3}
& IEEE 1901.2 & Not a European Harmonized Standard \\
& 148.5--500~kHz & \\\hline
USA & 47 CFR  \S15 & Rules for power line carrier \\
& 9--490/500~kHz& or carrier current systems apply\\
\hline
Japan & ARIB STD T-84& CSMA/CA required \\
&  10--450~kHz & \\\hline\hline
\multicolumn{3}{|c|}{Broadband PLC}\\\hline
Europe & EN~50561-1& Dynamic power control\\
& 1.6065--30~MHz & Static and dynamic notching\\
\hline
USA & 47 CFR  \S15 & Subpart~G for access BB PLC\\
&1.705--80~MHz & Interference mitigation and avoidance\\
&& Excluded bands and zones\\
\hline
\end{tabularx}
\end{center}
\end{table}

\subsubsection{BB PLC}
\label{sec:bbplcreg}

For BB PLC, radiated emissions become a bigger concern due to the higher signal frequencies and the asymmetries in power line networks. In Europe, the specification of harmonized emission limits has been complicated by the fact that relevant standards differentiate between a mains port and a telecommunication port of the tested equipment, and thus do not account for an intentional symmetric signal transmitted via the mains port. Therefore, it has been argued that measurement methods and emission limits need to be adjusted for PLC devices, see \cite{LL-Bookcollection:PLC-Ferreira-2010Chap3,LL-Bookcollection:PLC-Ferreira-2010Chap7}, \cite[Ch.~3]{lampe2015power}. This has been resolved relatively recently with the approval of EN~50561-1 \cite{LL-EN5056112012} in November 2012. This standard applies to in-home PLC systems operating in the 1.6--30~MHz band and it differentiates between a power port (only for power supply), a telecommunications/network port (only for communication signals) and a PLC port (for communication and power supply). It specifies maximal voltage levels for the PLC signals and the corresponding measurement procedures. The standard also requires dynamic power control and static and adaptive notching of frequency bands, which renders BB PLC systems cognitive ``radios'' (which was anticipated by the research community, e.g. \cite{praho2010cognitive,oh2009cognitive}).  Further standards, namely EN~50561-2 for access networks and EN~50561-3 for frequencies above 30~MHz, are under development.

The 47 CFR \S15 by the U.S.\ FCC \cite{LL-FCC08} defines in-house and access ``broadband over power line'' (BPL) systems. The former fall under the regulations for carrier current systems mentioned above. The latter are specifically addressed in Subpart~G for the band 1.705--80~MHz. Subpart~G sets out radiated emission limits, differentiating between medium-voltage (MV) and low-voltage (LV) installations, interference mitigation and avoidance methods including adaptive power control and frequency notching, administrative requirements which include registration of deployments in a database, and excluded frequency bands for overhead deployments and geographic exclusion zones. As in EN~50561-1, power adaptation and notching is intended to avoid harmful interference to radio services.  

The limits specified in EN~50561-1 and 47 CFR \S15 for BB PLC can be translated into power spectral density masks for a given termination impedance, e.g. \cite[Fig.~6.3]{LL-Bookcollection:PLC-Berger-2014-WholeBook}. This leads to a power spectral density (PSD) of about $-55$~dBm/Hz for transmission up to 30~MHz at an impedance of 100~$\Omega$, which is consistent with the PSD specifications in the standards IEEE~1901 \cite{IEEE1901} and ITU-T G.9964 \cite{LL-ITU9964}.

\subsection{PLC Standardization}
\label{sec:PLCstandardization}
There are numerous industry specifications for NB PLC systems that support link rates of up to a few kbps and operate in the application space of home and industry automation and for utility applications, e.g. see \cite[Ch.~7]{lampe2015power}, \cite[Ch.~2.2]{LL_Book:IoT-2011}, \cite{LL-dzung:2011}. Several of these were adopted as international standards in the late 1990s and early 2000s, including ISO/IEC 14908-3 (LonWorks), ISO/IEC 14543-3-5 (KNX), and IEC 61334-5-1/2/4. These systems have established a track record for PLC as a proven technology  for low data-rate applications. %, \cite[Table 2.1]{LL_Book:IoT-2011}, \cite[Fig.~10.1]{LL-Bookcollection:PLC-Berger-2014Chap10}. 
In the following, we first review the standards developement for BB PLC and then NB PLC standards for relatively higher data rates.

\subsubsection{BB PLC}

\begin{figure}[t]
\centering
\includegraphics[width=\columnwidth]{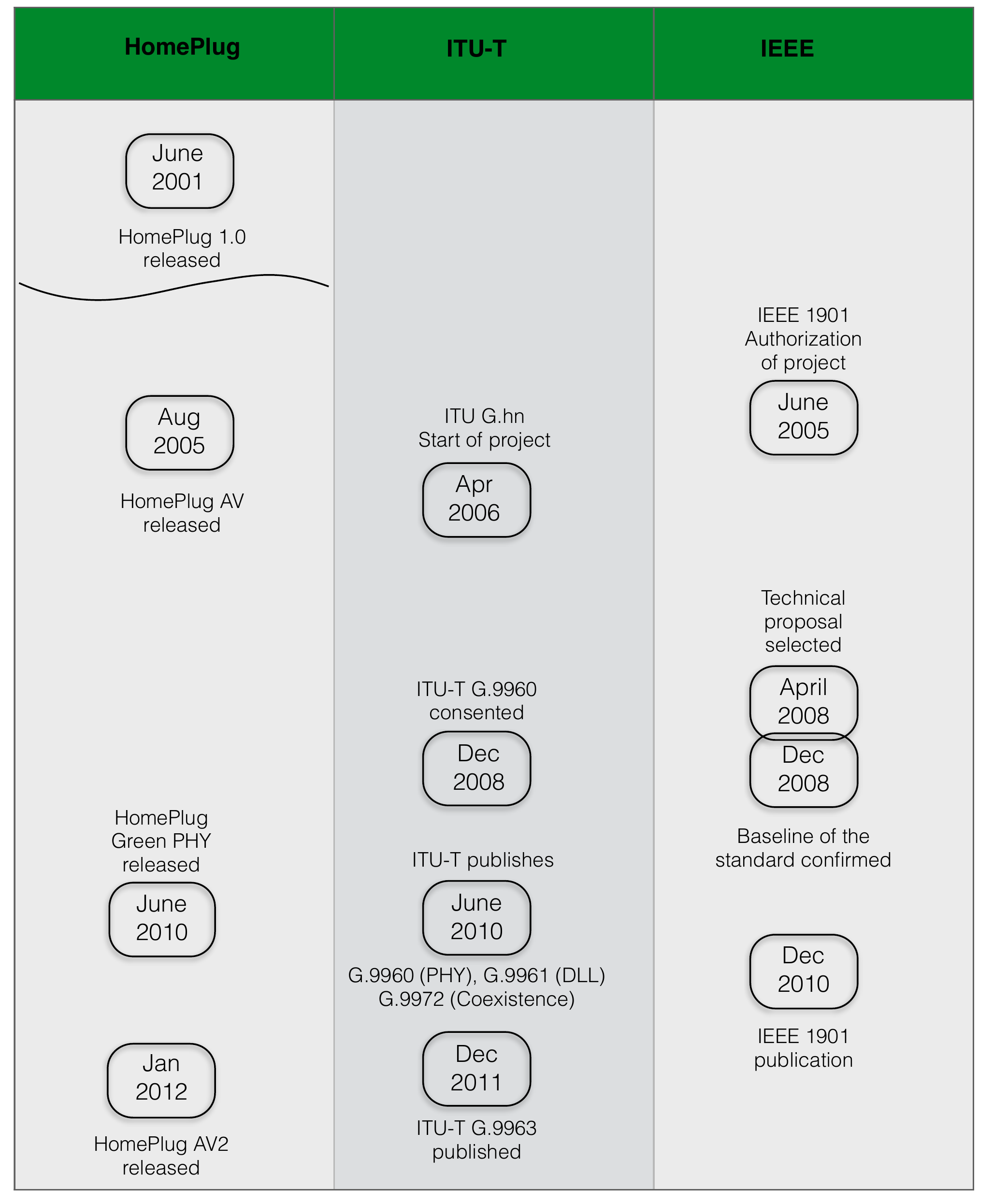}
\caption{Overview and timeline for the development of BB PLC specifications and standards.}
\label{f:BBstandards}
\end{figure}
The late 1990s saw a spur of activities in the PLC community developing BB PLC solutions for the access and in-home domains, eventually targeting data rates of hundreds of Mbps \cite{LL-commag1:2003,LL-commag2:2003}. This resulted in several industry specifications, mainly those backed by the HomePlug Powerline Alliance (HomePlug), the Universal Powerline Alliance (UPA) and the High-Definition Power Line Communication (HD-PLC) alliance. The left column of Fig.~\ref{f:BBstandards} shows some important steps in the evolution of HomePlug specifications, starting with HomePlug~1.0 released in June 2001 \cite{HomeplugStd,LL-lee:2003}, which was adopted by the Telecommunications Industry Association (TIA) as the international standard TIA-1113 in 2008. However,  the existence of different non-interoperable specifications has not been ideal for broad market success. Against this background, the consolidation of BB PLC systems in international standards was started by the IEEE P1901 Corporate Standards Working Group 
in June 2005 and the ITU-T standardization project G.hn in April 2006 \cite{LL-galli:2008,LL-oksman:2009}. In 2010, this resulted in the publication of the IEEE~1901 \cite{IEEE1901}  and the ITU-T G.9960/61 \cite{LL-ITU9960,LL-ITU9961} standards, which specify the physical and data link layers as well as coexistence mechanisms and PSD masks, see columns~2 and~3 of Fig.~\ref{f:BBstandards}. 

The IEEE standard uses the 2--30~MHz frequency band with an optional extended band of up to 50~MHz. It includes two multicarrier physical (PHY) layers, which are commonly referred to as OFDM via Fast Fourier Transform (FFT-OFDM) and Wavelet OFDM, respectively. The former is classic (windowed) OFDM, while the latter is a discrete wavelet multitone (DWMT) modulation \cite{LL-sun:2002}. They are non-interoperable but their coexistence is enabled by an inter-PHY protocol (IPP) \cite{GalliCOEX}, which later was extended to allow coexistence also with  G.9960, and this extension was called inter-system protocol (ISP). The FFT-PHY applies a Turbo code for forward error correction (FEC), while the Wavelet-PHY uses concatenated Reed-Solomon (RS) and convolutional codes, which can optionally be replaced by low-density parity-check (LDPC) convolutional codes. The PHY layers support multiple signal constellations and spectral masking as required by regulations discussed in Section~\ref{sec:bbplcreg}. On top of these two physical layers 
resides, via PHY layer convergence protocols, a common medium-access control layer that enables both CSMA and time-division multiple access (TDMA). 

While IEEE~1901 has provisions for in-home and access networks, ITU-T G.hn applies specifically to home networking. It does not apply only to PLC but also to communication over phone lines and coaxial cables. For PLC, it includes three bandplans, from 2~MHz to 25, 50, and 100~MHz, respectively. The spectral mask to comply with the emission limits outlined in Section~\ref{sec:bbplcreg}  is consistent with that used in IEEE~1901. Also, as in IEEE~1901, windowed OFDM with flexible bit loading is applied, and CSMA and TDMA are used for medium access. In contrast to the IEEE~1901 PHY modes, the physical layer of ITU-T G.hn uses LDPC block codes. The IEEE~1901 and ITU-T G.hn standards are non-interoperable. However, coexistence is enabled through the ISP specified in IEEE~1901 and ITU-T G.9972 \cite{LL-ITU9972}, whose support is mandatory for IEEE~1901 devices. 

En-route towards Gbps transmission, a multiple-input multiple-output (MIMO) transmission  extension to G.hn has been specified as ITU-T G.9963 \cite{LL-ITU9963}. Similarly, HomePlug published the HomePlug~AV2 standard \cite{LL-yonge:2013}, which is backward compatible with HomePlug AV and IEEE 1901 and includes MIMO transmission as well as efficient notching and power back-off to reduce emissions. At the other end of the data-rate spectrum, ITU-T G.9960 includes a low-complexity profile for reduced component cost and power consumption targeting the smart grid market. Similarly, the HomePlug Green PHY specification \cite{LL-HPGP} has been developed as a subset of the HomePlug AV standard for low power consumption and low cost, targeting the home-area network domain of smart grids. Fig.~\ref{f:BBstandards} summarizes the mentioned standards along the timeline of their publication dates. 

\subsubsection{NB PLC}

\begin{figure}[t]
\centering
\includegraphics[width=\columnwidth]{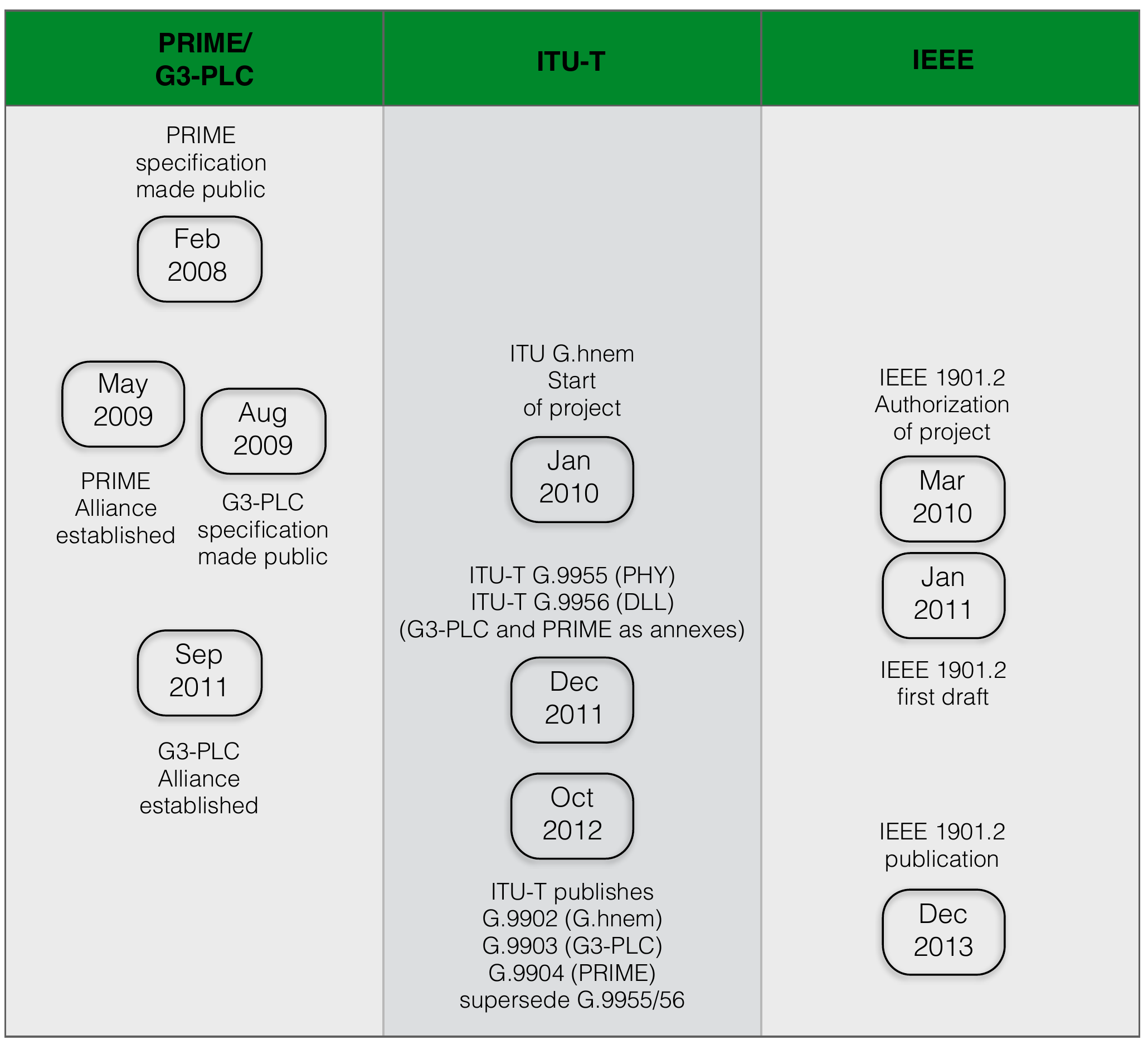}
\caption{Overview and timeline for the development of HDR NB PLC specifications and standards.}
\label{f:HDRstandards}
\end{figure}

\begin{figure}[t]
\centering
\includegraphics[width=\columnwidth]{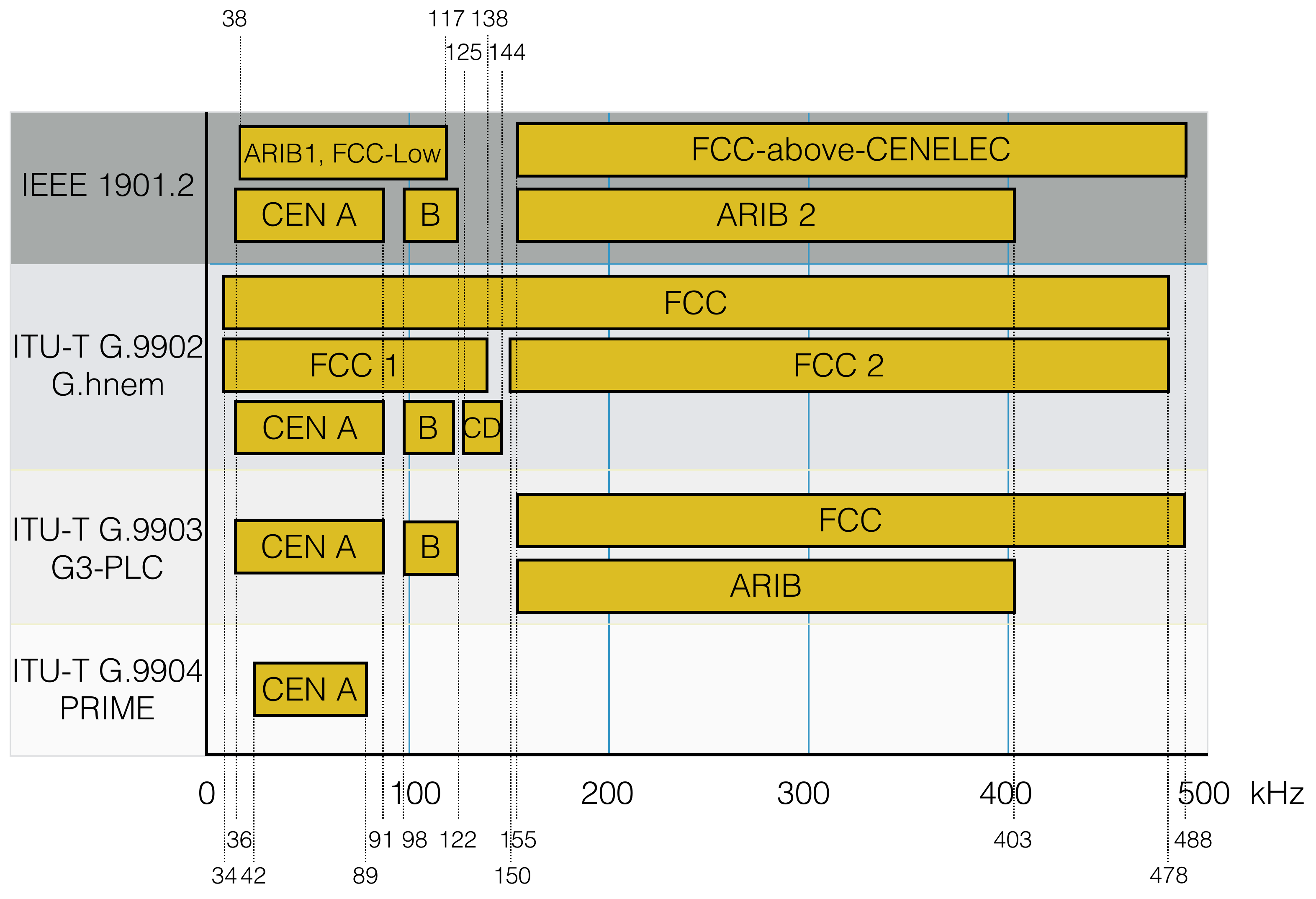}
\caption[]{Frequency bandplans for standards and specifications of HDR NB PLC systems\footnotemark{}, following bands available in different regions of the world (see Section~\ref{sec:nbplcreg}). The numbers are the center frequencies of the start and end tone for each of the bands rounded to the next kHz-integer value. Adapted from \cite{LL-galli:2015}.}
\label{f:HDRbands}
\end{figure}

The development and standardization of BB PLC systems have been followed by a wave of activity to specify NB PLC solutions for relatively high data-rate (HDR) transmission. These efforts have been driven by the demands for an effective smart grid communication infrastructure \cite{LL-lampe:2011,Bookcollection:Vision-SmartGrid}. Fig.~\ref{f:HDRstandards} provides an overview of the development of major industry specifications and SDO standards for such HDR NB PLC systems. ``HDR''  means that data rates of tens to hundreds of kbps are achieved using the 3--500~kHz frequency band. In particular, in accordance with the frequency bands available in different regions of the world as described in Section~\ref{sec:nbplcreg}, the specifications listed in Fig.~\ref{f:HDRstandards} have defined different bandplans as shown in Fig.~\ref{f:HDRbands}. 
\footnotetext{The PRIME v1.4 PHY specification extends the usable frequency band  to 42--472~kHz.}

The industry specifications PRIME (Powerline Related Intelligent Metering Evolution) and G3-PLC have been developed by the PRIME Alliance and the G3-PLC Alliance respectively. Following the BB PLC example, multicarrier modulation and in particular OFDM has been adopted for the PHY layer. In contrast to BB PLC, differential modulation is mandatory in both standards, but support of coherent modulation was added to G3-PLC later on. Differential modulation avoids the need for channel estimation and is thus better suited for transmission of shorter messages and more robust to channel variations. Further emphasizing simplicity, convolutional coding is used. In case of PRIME, even this is optional, while G3-PLC adds an outer RS code.  G3-PLC also specifies a robust mode that uses additional repetition. Such a mode has been added to the latest version of PRIME. 

In 2011, the ITU-T published recommendations ITU-T G.9955 for the PHY layer and ITU-T G.9956 for the link layer, which included PRIME and G3-PLC as well as the new G.hnem technology. The latter uses coherent transmission. This has been reorganized into standards ITU G.9902-04, as shown in Fig.~\ref{f:HDRstandards}. ITU-T G.9904 (2012) adopted PRIME v1.3.6 as is, whereas G3-PLC adopted in ITU-T G.9903 evolved since its first submission to ITU-T and went through three major revisions (2012/2013/2014). In 2013, the IEEE published the IEEE 1901.2 standard, which is based on G3-PLC. However, as outlined in \cite{LL-galli:2015}, IEEE 1901.2 and ITU-T G.9903 have differences that render them non-interoperable. But IEEE 1901.2 includes a NB-PLC coexistence protocol that has also been adopted in ITU-T G.9903 (2014), which enables devices using these standards to coexist.

%{\color{red}XXX Add something about MAC? Add something about L2 routing, adaptation layers, and coexistence? }

\subsection{Industrial Solutions}

Examples from each class of PLC, namely UNB, NB and BB, have been implemented in products and find different applications. Example of UNB technologies include the Turtle system from Landis+Gyr and two way automatic communications system (TWACS) from Acalara supporting data rates from sub 1 bits/sec to 10’ of bits/sec while reaching distances of 150 km. For NB, several chip vendors including Renesas, STM, Maxim, Texas Instruments, SemiTech, Semtech support PRIME, G3, IEEE 1901.2 standards. Challenges in developing solutions for these standards include building high performance modems that can handle PLC interference, interface to low impedance lines, challenge at the MAC level and building reliable mesh networks. Large scale deployments of these technologies have been announced in France and Spain. BB-PLC vendors include Broadcom and Qualcomm implementing Homeplug AV2 standard for in-home applications with MIMO support. In a different application area, for wireless power transfer/charging and communications, 
several products have been announced for the Qi, A4WP wireless charging standards.

\section{PLC Medium}\label{sec:plc_medium}
The power distribution network was not conceived as a medium for data transmission. As a medium it has peculiar characteristics in the frequency band of interest, i.e. above 10 kHz and up to 300 MHz. The primary characteristics are the high frequency selectivity and attenuation: these are due to multipath signal propagation caused by the presence of multiple branches (discontinuities), unmatched loads and high frequency selective low impedance loads. Time variations are also exhibited when the network topology changes and/or the loads change. Furthermore, the PLC medium experiences high levels of noise injected by devices connected to the power grid or coupled through electromagnetic phenomena. In the following, the PLC channel in home and outdoor networks are described. The single-input single-output (SISO) channel, the multiple user (MU) and the multiple-input multiple-output (MIMO) channels are considered individually in the following to highlight the main distinctive properties.

\subsection{SISO Channel}
In this section, the channel properties are assessed in terms of the main and most commonly used statistical metrics, namely the average channel gain (ACG), the root-mean-square delay spread (RMS-DS) and the coherence bandwidth (CB). Furthermore, data from measurements made in different countries are compared.

\begin{figure}[tb]
\centering
\includegraphics[width=\figurewidth]{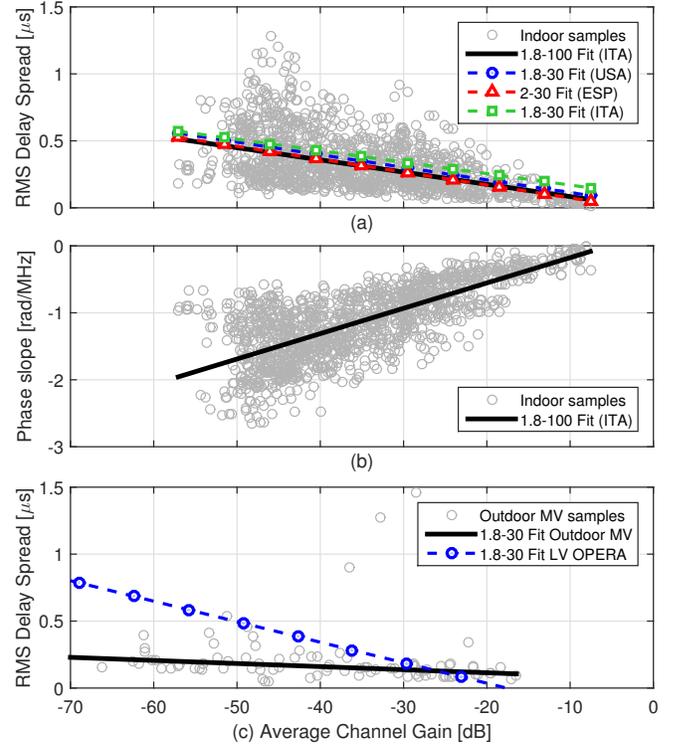}
\caption{RMS-DS versus ACG for the in-home scenario in the 1.8--100 MHz band (a) and the outdoor MV scenario in the 1.8--30 MHz band (c), with the corresponding robust regression fit. The robust fits for the 2--30 MHz Spanish (ESP) \cite{175}, 1.8--30 MHz USA \cite{97} and our (ITA) 1.8--30 MHz measurements (a), and the 1.8--30 MHz LV OPERA \cite{OperaD4} model (c) are also shown. The unwrapped phase slope versus ACG for the in-home scenario, together with the robust fit, are reported in (b).}
\label{fig:Statistics}
\end{figure}

The in-home scenario is considered first. In Fig.~\ref{fig:Statistics}a and Fig.~\ref{fig:Statistics}b the RMS-DS and the unwrapped phase versus the ACG (in dB scale) respectively are reported. The circles correspond to the scatter plot of all measured values in a campaign conducted in Italy (described in \cite{TCOM1}) in the band 1.8--100 MHz, while the lines correspond to the robust regression fit. Looking at Fig.~\ref{fig:Statistics}a, note that the RMS-DS and the ACG are negatively related \cite{Galli_09}. This indicates that the channel attenuation is due to multipath propagation and it increases for highly dispersive channels. The figure reports also the robust fit of the measurements carried out in Spain and in the USA, described in \cite{175} and \cite{97} respectively. Despite the different wiring practices, the robust fit curves are very similar. 

Fig.~\ref{fig:Statistics}b shows the relation between the phase slope of the robust fit of the unwrapped phase of the channel frequency response (CFR) and the ACG. The phase slope offers some information about the average delay introduced by the channel and thus, in turn, on the length of the backbone, i.e. the shortest electrical path between the transmitter and the receiver. The greater the magnitude of the phase slope, the larger the expected wire length and number of branches. Consequently, the higher the attenuation, the lower the ACG, since the channel attenuation increases with the distance and the number of branches connected to the backbone.

Another important aspect is the definition of the channel. In contrast to the wireless scenario, there is no expectation that the PLC channel is Rayleigh distributed. The amplitude of the channel frequency response is well fitted by the log-normal distribution, as first reported in \cite{Galli_09} and then in \cite{97}. However, this is scenario dependent and deviations in the distribution tails can be encountered \cite{TCOM1}. Furthermore, a correlation is manifested between the channel response at different frequencies \cite{IEICE_14}. Finally, the channel response can exhibit a periodically time-variant behavior as a result of the periodic variations, with the mains AC voltage, of the load impedance \cite{Canete_06}. This is particularly true at frequencies below 2 MHz.     

Now, let us turn attention to the outdoor PLC channel. In Fig.~\ref{fig:Statistics}c, the outdoor MV channel from measurements conducted in Italy \cite{SGC_14} is considered. The robust fit for the outdoor LV channel, from the EU OPERA project measurements \cite{OperaD4}, is also depicted. In particular, note that the RMS-DS robust fit slope of the outdoor MV channels is approximately half the slope of the in-home channels considered in Fig.~\ref{fig:Statistics}a. This is because the MV channels are more attenuated due to longer cables and, furthermore, they exhibit lower RMS-DS due to reduced multipath propagation into a network topology that has fewer branches.  Contrariwise, the slope of the OPERA LV channels is almost double the slope of the in-home channels. This is due to the large number of signal reflections introduced by the typical network structure that consists of a backbone with many short branches connecting premises. The high attenuation in the OPERA LV channels may be explained 
by the resistive characteristics of the deployed cables.

The channel at lower frequencies, e.g. in the NB spectrum of 9--500 kHz, is less attenuated than the channel at frequencies beyond 2 MHz, i.e. the BB channel both of the indoor and in the outdoor environments. This is shown in Table~\ref{tab:StatMetrics} where the average ACG, RMS-DS and CB are reported for different bands and scenarios. Data were obtained from \cite{TCOM1}, \cite{SGC_14}, \cite{OperaD4}, \cite{Nassar2012} (NB-PLC measurements in Indian and Chinese sites), \cite{Cortes2015} (for the CENELEC-A band of 3--95 kHz). The NB channel characterization has been less documented than the BB one, especially for the outdoor scenario. Given the relevance of recently developed NB PLC technology, it would be beneficial to further investigate the NB channel both in indoor and outdoor scenarios and to report a detailed analysis.

\begin{table}[th]
	\centering
	\caption{Average statistical metrics for different channel scenarios in different frequency bands.}
	\begin{tabular}{c | c | c | c | c}
       Scenario & Band & ACG & RMS-DS & CB \\
       &  & (dB) & ($\mu$s) & (kHz) \\
       \hline
       In-Home & 1.8--100 MHz & $-35.41$ & $0.337$ & $288.11$ \\
       In-Home & 1.8--30 MHz & $-31.91$ & $0.394$ & $216.48$ \\
       Outdoor MV & 1.8--30 MHz & $-40.53$ & $0.491$ & $458.58$ \\
       OPERA LV & 1.8--30 MHz & $-54.64$ & $0.581$ & $140.63$ \\
       \hline
       OPERA LV & 9--500 kHz & $-32.02$ & $2.345$ & $30.69$ \\
       Outdoor LV \cite{Cortes2015} & 3--95 kHz & $\sim-35$ & $\sim19$ & $\sim4$ \\
       Outdoor LV \cite{Nassar2012} & 3--500 kHz & $-$($15$--$33$) & $2.2$--$4$ & - \\
	\end{tabular}
	\label{tab:StatMetrics}
\end{table}

\subsection{Multiuser Channel}
When we consider a network of nodes connected to the same power grid, it becomes important to characterize the multiple user (MU) channel. In this respect, the underlying network structure deeply affects the channel properties, and in turn the achievable MU communication performance.

The MU PLC network has, in general, a tree structure, so that pairs of nodes share part of the wireline network. For instance, if we consider a pair of channels from a given transmitter to two distinct receivers, they share part of the communication link up to a certain node (named the pinhole or keyhole) where branches then depart towards the final receiving destinations. This structure gives rise to a phenomenon known as the keyhole effect, which was documented in the context of cooperative multi-hop PLC in \cite{Lampe_12} and, later, considering the physical layer security in PLC networks in \cite{ISPLC13}. It is important to note that the MU concept holds for both the indoor and the outdoor scenarios, and for any implemented transmission scheme, such as SISO or MIMO. However, within this section, the presented results are based on the 1300 SISO in-home channel measurements discussed in \cite{TCOM1}.

\begin{figure}[tb]
\centering
\includegraphics[width=\figurewidth]{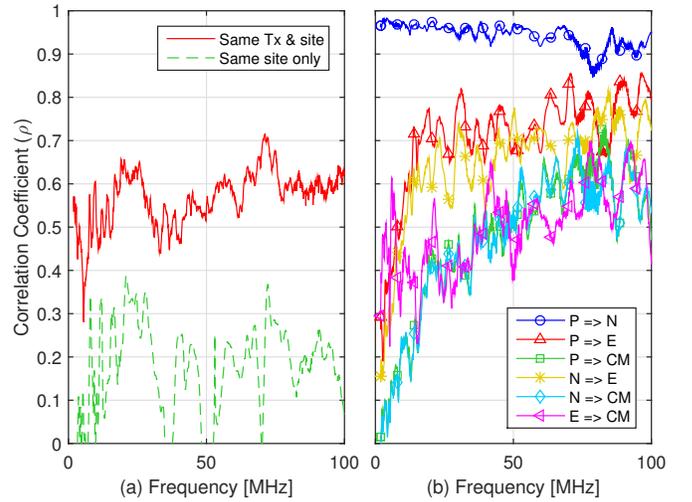}
\caption{MU correlation coefficient between SISO PLC channels sharing, or not, the same transmitter (a), and spatial correlation coefficient among all the star-style receiving mode combinations for the MIMO channels (b).}
\label{fig:UsersMIMOcorr}
\end{figure}

A key aspect of the MU channel is that the communication links are correlated or, in other words, there is a certain level of determinism. This is significantly different from the wireless, where MU channel diversity is often introduced by rich scattering (multipath) propagation. Now, to quantitatively show this, we can compute the spatial correlation coefficient $\rho$, which is defined as $\rho=E[H^{(i)}(f){H^{(j)}}^*(f)]$ for pairs of distinct $i,j$ channels. In Fig.~\ref{fig:UsersMIMOcorr}a, $\rho$ is reported for channels of a given site sharing the same transmitter as well as when the constraint of having the same transmitter is removed. The data base of measurements in \cite{TCOM1} is used. 
It should be noted as $\rho$ takes high values, approximately equal to 0.5, along almost the entire frequency range for channels sharing the same transmitter. This high spatial correlation reduces the available channel diversity. 

Spatial correlation is exhibited also in the MIMO channel, as discussed in the next section.

\subsection{MIMO Channel}
MIMO systems are popular in the wireless domain where they deploy multiple transmitting and receiving antennas. Also in the PLC context, MIMO transmission can be established by exploiting the presence of multiple conductors. In home networks, for instance, the power network comprises three wires: the phase (P), the neutral (N) and the protective earth (E) wires\footnote{The protective earth acts as a return path for the power supply in the case of an insulation fault.}.
\begin{figure}[tb]
\centering
\includegraphics[width=\figurewidth]{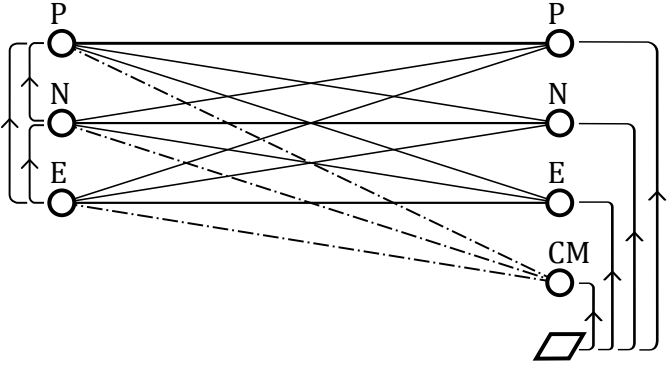}
\caption{Possible MIMO transmission modes in a PLC network, according to the STF-410.}
\label{fig:MIMOnetwork}
\end{figure}
%If a reference plane is used, a MIMO PLC system can exploit not only the differential propagation mode signals, but also the common mode (CM) signal \cite{ETSI1,Schwarz}. This configuration is schematically depicted in Fig.~\ref{fig:MIMOnetwork}.

At the transmitter side, the differential voltage signal can be injected between pairs of wires giving rise to three different signals, referred to as delta $\Delta$ modes. Due to the Kirchhoff's laws, only two $\Delta$ signals can be injected at the same time. Contrariwise, if a reference plane is used at the receiver side, the signal can be observed between one conductor and the reference plane. This configuration is referred to as star-mode ($S$-mode). The number of possible $S$-modes is three, corresponding to the number of wires. However, an additional mode, named common mode (CM), can be exploited \cite{ETSI1,MIMO_PLCimpl}. Large TV screens, for instance, include a large metal plane that can act as a reference plane. If the reference plane is not available, then, the delta reception mode can be used. The CM flows with the same intensity and direction through the P, N and E wires. For EMI reasons, the CM is used only at the receiver side. Hence, as shown by Fig.~\ref{fig:MIMOnetwork}, a $2\times4$ MIMO 
transmission can be established. However, it should be noted that, due to Kirchhoff's laws, the fourth signal collected by the start-style coupler is linearly dependent on the other three signals, but it can provide a signal-to-noise ratio gain as well as a capacity gain \cite[Ch. 1 and 5]{LL-Bookcollection:PLC-Berger-2014-WholeBook}.

In the MIMO scenario, the PLC channels are correlated due to the symmetry and determinism of the wiring structure. The correlation of the channel responses among different $S$-style receiving modes (where e.g. P$\Rightarrow$N indicates the correlation among phase and neutral) is reported in Fig.~\ref{fig:UsersMIMOcorr}b. The figure shows that the correlation coefficient is even higher than that found in the MU channel sharing the same transmitter in Fig.~\ref{fig:UsersMIMOcorr}a. In particular, high levels of correlation are exhibited among the channels P$\Rightarrow$E and N$\Rightarrow$E, with the highest values experimented among the P$\Rightarrow$N channels responses, especially for lower frequencies. These high correlation values are due to the fact that the power is delivered through phase and the neutral wires, which are positioned one next to the other and follow the same path from the transmitter to the receiver. 
%Contrariwise, although the protective earth channel follows almost the same path, it comprises the device chassis (for grounding purposes in case of faults) as well as the wires for an overall grounding to the physical earth. This is the reason underlying the slightly lower values of correlation among P$\Rightarrow$E and N$\Rightarrow$E.

Not only the MIMO channel responses are correlated but also the noise. Colored and spatially correlated noise are considered in \cite{Pittolo_14} for the ${2\times 4}$ scheme (see also Section~\ref{sec:NoiseProperties}). 

A first study of the performance improvement achieved through the use of a ${2\times 4}$ MIMO communication method w.r.t. the SISO system, in the 2--30 MHz frequency band, is reported in \cite{MIMOforPLC}. Additive white Gaussian noise (AWGN), spatial multiplexing (beamforming at the transmitter) and zero forcing (ZF) detection are assumed. Precoded spatial multiplexing is analyzed in \cite{SptMuxPLC}, while \cite{MIMO_PLCimpl} describes implementation results. Finally, the performance improvements with the exploitation of a $2\times2$ MIMO scheme with differential signal transmission and reception are discussed in \cite{74}.

\subsection{Channel Response Modeling}
Modeling the PLC channel impulse and frequency response has attracted considerable attention. Models can be categorized as top-down when the approach is phenomenological, and as bottom-up when a physical description of signal propagation using transmission line theory is used. Furthermore, the model can be deterministic or statistical (random).   

A deterministic top-down multipath propagation model was firstly proposed in \cite{6} targeting the BB frequency spectrum. Then, it was improved in \cite{21}, extended in statistical terms firstly in \cite{Tonello2007} and refined in \cite{155}. Other top-down BB random channel models were proposed in \cite{57} (in frequency domain) and \cite{97} (in time domain). Inspired by \cite{Tonello2007}, a MIMO statistical top-down model was presented in \cite{Hashmat_11} for the BB spectrum.

First attempts for bottom-up NB modeling were presented in \cite{19} for the in-home scenario, while in \cite{196} for the NB outdoor low-voltage scenario. More recently, bottom-up models based on the $s$-parameters and ABCD-matrix representations were proposed in \cite{12,23,10}. While \cite{12} and \cite{10} consider the BB frequency range, \cite{23} considers the NB spectrum providing a validation of the model in time domain. In what follows, all the referred works consider the BB spectrum. In particular, it was shown that a random extension of the bottom-up model is possible by using a random (although simplistic) topology representation in \cite{5}. This bottom-up modeling approach can also be exploited to include the periodic time variant channel changes by adding the time variant behavior of load impedances \cite{98}. A statistically representative random topology model for home networks was presented in \cite{60} together with an efficient way to compute the channel transfer function in complex 
networks referred to as voltage-ratio approach. This model was used to infer the BB channel statistics as a function of the network geometry in \cite{TonelloPartII}.

While bottom-up modeling offers a tight connection with physical propagation of PLC signals in a certain network, top-down modeling is particularly attractive for its low complexity. It is therefore foreseen that refined top-down models will be developed in the future. Recently, it has been shown that the simplest way to model the SISO CFR is to directly generate the amplitude and phase as a vector of correlated complex random variables, whose marginals have log-normal amplitude and uniform phase distribution \cite{IEICE_14}. An analytic expression for the correlation matrix in the frequency domain (i.e. between the channel samples in frequency) can be derived from experimental measures. Finally, by exploiting the relation existing between the Pearson (linear) and the Spearman (rank) correlation, the multivariate CFR distribution can be generated.

\subsection{Line Impedance}
\begin{figure}[tb]
\centering
\includegraphics[width=\figurewidth]{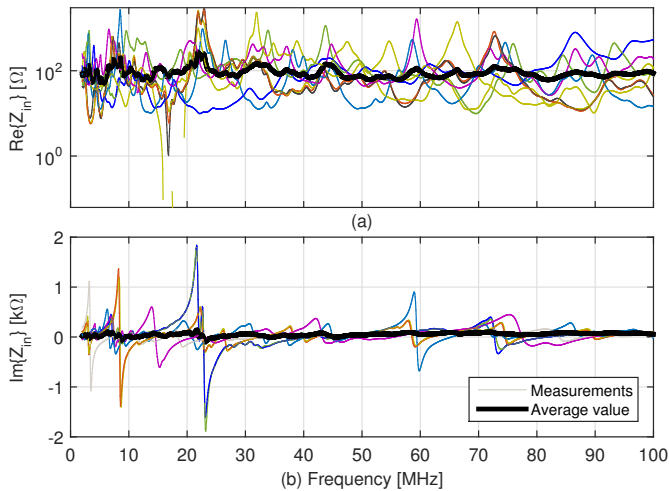}
\caption{Real part (a) and imaginary part (b) of the input line impedance for the in-home scenario in the broadband frequency range 1.8--100 MHz.}
\label{fig:Impedance}
\end{figure}

Not only the channel response is important, but also the line impedance has to be considered since it affects the design of the analog front-end of the PLC modem. A low line impedance at the transmitter port makes the injection of the voltage signal challenging. The measurement results have shown that the line impedance can be significantly low (in the order of few ohms) especially at low frequencies. For instance, in the access network, this is due to the fact that the home network acts as many parallel loads attached to the access port \cite{ferreira1999power}. In the home network, the line impedance (at the outlets) exhibits a highly frequency-dependent behavior, as shown in Fig.~\ref{fig:Impedance}. Interestingly, it is mostly inductive and the real part increases at high frequency. This behavior can be especially noted looking at the single realizations in Fig.~\ref{fig:Impedance}a (each one represented with a different color). Therefore, it is expected that broadband PLC can be less affected by this 
issue than NB PLC and the broad spectrum provides channel and impedance diversity which can simplify the design of impedance adaptation techniques \cite{Antoniali2013}. 

\subsection{Noise Properties and Models}\label{sec:NoiseProperties}
The PLC medium is affected by severe noise with stationary (typically referred to as background) and nonstationary (referred to as impulsive) components, which depend on the considered application scenario (e.g. indoor or outdoor and LV or MV). The former is a combination of conducted noise and coupled radio signal contributions. The latter can be cyclostationary, with a repetition rate that is equal to, or double, that of the mains period, or bursty and cyclostationary, with a repetition rate that is high between 50 and 200 kHz, or aperiodic. The first two components are referred to as periodic noise synchronous and asynchronous with the mains frequency \cite{134}. The periodic synchronous noise originates from silicon controlled rectifiers (SCR) in devices, while the asynchronous noise is due to the switching activity of power supplies. The characterization in the time and frequency domain of PLC noise can be done by observing it at the receiver port \cite{32} or at the source \cite{TPD_15}. The aperiodic 
noise is the most unpredictable component and it is due to the connection and disconnection of appliances from the power delivery network. The amplitude of the aperiodic noise can be significantly larger than that of the other impulsive noise components. Beside the amplitude, the aperiodic impulsive noise is typically described by the duration and the inter-arrival time \cite{148}. The statistics of these quantities depend on how the impulsive noise events are identified and measured.  

The PLC background noise is usually modeled with a stationary Gaussian colored process having a frequency decreasing power spectral density (PSD) profile, according to a polynomial \cite{5} or exponential \cite{74} function of frequency. Typical noise PSD trends, having different floor levels and profiles, have been reported in \cite{IEICE_14} for the in-home, the outdoor LV and MV scenarios. The main differences are related to the network structure and topology, as well as to the type of connected loads. For example, indoor networks are characterized by many loads interconnected through a grid deploying short cables. This prevents the noise attenuation, leading to high levels of noise at the receiver side.  Outdoor networks, instead, are affected by the noise contribution generated by the overall industrial and residential consumers, by the inverters used in renewable generation plants, as well as by the RF interference coupling into the grid.

A model for the periodic noise terms that is based on a deseasonalized autoregressive moving average is presented in \cite{139}. Several authors suggest to fit the amplitude of the noise in the time-domain with the Middleton's class A distribution, e.g., \cite{147}. In \cite{140} it was speculated that the Nakagami-$m$ distribution is more appropriate. However, the results in \cite{140} are obtained from too few measurements to be considered conclusive. Some further studies reveal that the normal assumption on the noise statistics holds true if the periodic time-variant nature of the noise is accounted for \cite{146} and the impulsive noise contributions are removed from the measures \cite{135}. Based on this, filtering a stationary process with an LPTV system is proposed in \cite{Nassar_12} to model cylostationary Gaussian distributed noise. In \cite{134} a Markov-chain model is proposed to model the ensemble of components. In \cite{141} the noise at the source is modeled with a non-Gaussian distribution as 
in \cite{5} and then the noise at the receiver is obtained by filtering it with the channel generated with a top-down channel response generator. It is found that with a sufficiently large number of noise sources, the overall noise at the receiver approaches the Middleton's class A distribution. In \cite{142} another top-down channel generator is used, instead, to filter the source noise. 

\begin{figure}[tb]
\centering
\includegraphics[width=\figurewidth]{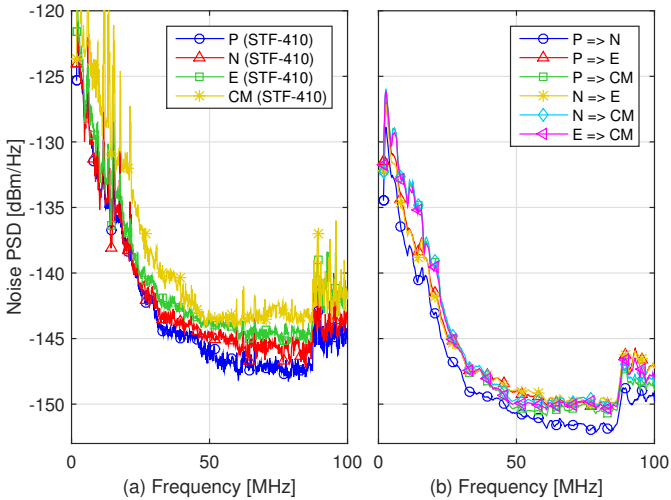}
\caption{Noise PSD profiles for the star-style receiving modes measured by the STF-410 (a) and the computed C-PSD among the different modes (b).}
\label{fig:NoiseETSI}
\end{figure}

Noise in the MIMO context has not been thoroughly studied yet. Experimental results in the home scenario are reported in \cite{ETSI3}, \cite{MIMOnoise_effect} and a model to account for the spatial correlation of noise is proposed in \cite{Pittolo_14}.  In this respect, Fig.~\ref{fig:NoiseETSI}a shows the noise PSD measured in the home scenario at the P, N, E and CM ports using an $S$-style coupler. While, Fig.~\ref{fig:NoiseETSI}b reports the cross-PSD (C-PSD) among each pair of noise modes. The C-PSD allows us to highlight the spatial cross-correlation that exists between the noise signals in frequency domain. It is defined as the covariance between the noise signals observed at modes $S_i$ and $S_j$ that is given by $R_{i,j}(f)=E[N_{S_i}(f)N_{S_j}^*(f)]$ with $i\neq j$, where $N_{S_i}(f)$ is the Fourier transform of the noise experienced by the mode $S_i$ \cite{Pittolo_14}.
In particular, from Fig.~\ref{fig:NoiseETSI}a note that the CM experiences the highest noise PSD, while the other three modes, i.e. P, N and E, have approximately the same PSD. The PSD significantly increases above 87 MHz due to coupled FM broadcasting radio signals. Fig.~\ref{fig:NoiseETSI}b, instead, shows that the C-PSD profiles resemble the PSD trends depicted in Fig.~\ref{fig:NoiseETSI}a, although they are lower. This is because the C-PSD elements are a linear combination of the PSD profiles in Fig.~\ref{fig:NoiseETSI}a and depend on a time-domain correlation coefficient that ranges between 0 and 1, as discussed in \cite{Pittolo_14}. The lowest noise C-PSD profile is exhibited between the modes P and the N since the corresponding PSD profiles are the lowest ones. Contrariwise, the combinations P$\Rightarrow$CM, N$\Rightarrow$CM and E$\Rightarrow$CM are affected by higher and similar noise levels. However, Fig.~\ref{fig:NoiseETSI}b shows that the noise C-PSD exhibits non-negligible levels and 
significant differences among the spatial modes.

\section{Physical Layer Performance}\label{sec:phy}
From the characterization of the medium, it is possible to assess the performance of the PHY layer. The Shannon capacity is the common metric used to determine the theoretical achievable rate limit. The secrecy capacity is another metric and it refers to the rate achievable by a communication that grants perfect confidentiality and secrecy. The results discussed in the first two sections, namely Section~\ref{ssec:capacity} and Section~\ref{ssec:sec_cap}, deal with the BB indoor scenario considering both SISO and MIMO transmission. Afterwards, PHY layer aspects and possible improvement directions are discussed considering both NB and BB systems in Section~\ref{ssec:PHYissues} and Section~\ref{ssec:PHYimprov}, respectively.

\subsection{Capacity}
\label{ssec:capacity}
The capacity depends on the channel, e.g., SISO or MIMO, the bandwidth and the noise assumptions. The true capacity of the PLC channel is unknown since a full characterization of noise properties and associated statistics has yet to be achieved. Typically, capacity is computed under the stationary Gaussian noise assumption, as we do in the following. We also assume that the transmitted signal has a PSD of $-50$ dBm/Hz up to 30 MHz and $-80$ dBm/Hz beyond 30 MHz (according to the HomePlug AV2 standard \cite{LL-yonge:2013}). 

\begin{figure}[tb]
\centering
\includegraphics[width=\figurewidth]{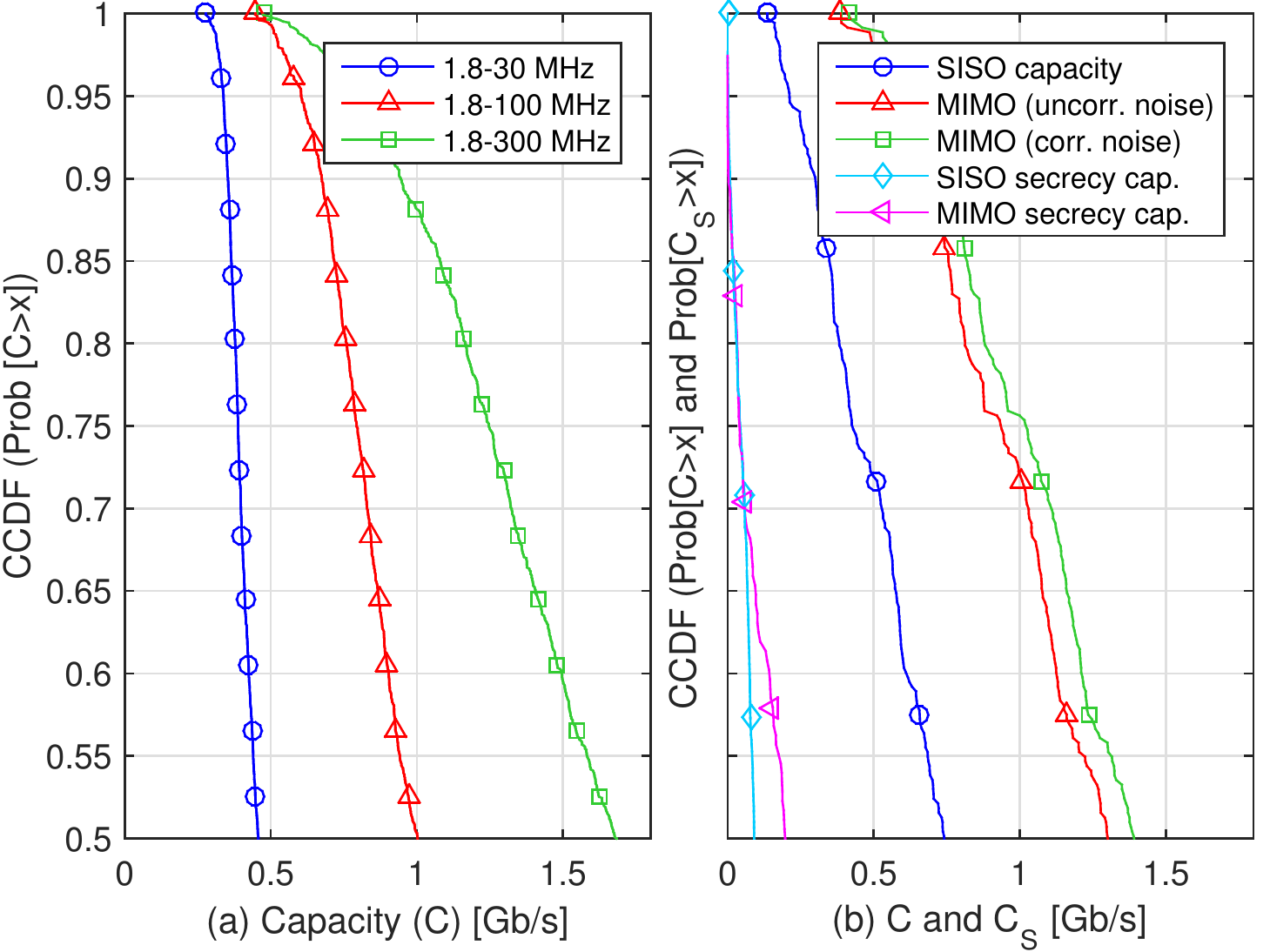}
\caption{Performance improvement due to bandwidth extension (a) and to the MIMO transmission in the 1.8--100 MHz band together with a colored and correlated Gaussian background noise assumption (b).}
\label{fig:BandMIMOperf}
\end{figure}

Firstly, we focus on the SISO in-home channel and on the gain attainable by a signal bandwidth extension. Fig.~\ref{fig:BandMIMOperf}a reports the capacity complementary cumulative distribution function (CCDF) with measured channel responses and typical background noise PSD (see \cite{TCOM1} for details). A bandwidth extension is beneficial, indeed. For 50\% of cases a rate exceeding 1.7 Gb/s can be achieved with a band 1.8--300 MHz, almost doubling the 1 Gb/s achieved in the 1.8--100 MHz band. 

Now, we turn our attention to the MIMO scenario by exploiting the channels measured by the ETSI special task force 410 (STF-410) discussed in \cite{ETSI1}. Herein, we also wish to show the effect of the noise when its time and space correlation is taken into account. The spatial correlation of the noise is considered according to the model described in \cite{Pittolo_14}. Fig.~\ref{fig:BandMIMOperf}b shows that MIMO provides significant gains over SISO with the same total PSD constraint. The difference between the SISO results in Fig.~\ref{fig:BandMIMOperf}a and Fig.~\ref{fig:BandMIMOperf}b are not pronounced and are attributable to the distinct CFR databases and slightly different background noise PSDs used for each experiment. The worst MIMO performance is obtained with the colored and spatially uncorrelated noise assumption, while colored and spatially correlated noise lead to a further capacity improvement if precoding is implemented. This is due to the fact that the knowledge of the noise correlation 
matrix facilitates its mitigation at the receiver side.

\subsection{Secrecy Capacity}
\label{ssec:sec_cap}
The secrecy capacity is another less common performance metric. It indicates the amount of information that can be reliably and securely exchanged between two nodes, without any disclosure of information towards a third party, or adversary, commonly referred to as eavesdropper or wiretapper. The secrecy capacity is studied in the context of so called physical layer security (PLS). While PLS has been studied extensively in wireless, the first study in PLC was reported in \cite{ISPLC13} and then extended in \cite{IET_14}. In the following, the main concepts concerning the PLS in PLC networks are briefly summarized.

From an information-theoretic point of view, a wiretap channel consists of a transmitter (Alice) that wants to send a confidential signal to a legitimate receiver (Bob) without any leakage of information towards a malicious eavesdropper (Eve), which tries to disclose the message. In the presence of Gaussian noise, the secrecy capacity under a power constraint is defined as $C_S=\max_{f_x\in\mathcal{F}}[I(x,y)-I(x,z)]^+$, where $x$ is the signal transmitted by Alice, while $y$ and $z$ are the signals received by Bob and Eve, respectively. The quantity $f_x$ denotes the probability density function (pdf) of $x$, whereas $\mathcal{F}$ is the set of all possible pdfs of the input signal $x$. The terms $I(x,y)$ and $I(x,z)$, instead, represent the mutual information among $x$ and $y$ or $z$, respectively. Moreover, $[q]^+=\max(q,0)$ so that $C_S$ is set to zero when Eve has a better channel realization than Bob. Since the mutual information terms are convex in $f_x$, a lower bound $R_S$ for the secrecy capacity 
can be formulated as $C_S\geq\left[\max_{f_x\in\mathcal{F}}[I(x,y)]-\max_{f_x\in\mathcal{F}}[I(x,z)]\right]^+=R_S$ \cite{Jorswieck_10}. In \cite{ISPLC13} it was proved that, differently from capacity, the secrecy capacity is upper bounded by a constant value even if the power indefinitely increases. Furthermore, it was shown in \cite{ISPLC13} and \cite{IET_14} that the channel statistics (log-normal in PLC and not Rayleigh as in wireless), as well as the spatial correlation of the channel (introduced by the keyhole effect), may further limit the secrecy capacity.

In Fig.~\ref{fig:BandMIMOperf}b, the secrecy capacity $C_S$ for the SISO and the MIMO transmission schemes is compared to the capacity $C$. The figure shows that the SISO secrecy capacity is considerably lower than the unconstrained capacity but, with the use of MIMO transmission, it can increase \cite{IET_14,6812346}. The secrecy capacity is in general low because it is upper bounded (as a function of power \cite{ISPLC13}) and it is obtained as the difference among the rates of the intended receiver and of the eavesdropper.

\subsection{Practical PHY Layer Issues: Modulation}
\label{ssec:PHYissues}
In order to achieve the channel capacity, advanced modulation and coding schemes have to be deployed. Current PLC technology deploys powerful channel coding schemes, such as concatenated Reed-Solomon codes, convolutional codes, turbo codes, low density parity check (LDPC) codes, or a combination of these techniques, together with high order modulation. Furthermore, to overcome burst of errors introduced by noise and channel frequency response notches, interleaving can be deployed in order to spread in time and in frequency (over the sub-channels in multicarrier modulation schemes) coded blocks of bits or symbols. At the moment, NB systems use simpler coding techniques e.g., convolutional codes with bit interleaving in ITU-T G.9904 (PRIME) and convolutional and Reed-Solomon codes in ITU-T G.9903 (G3-PLC) and in IEEE 1901.2. Contrariwise, BB systems provide high speed communication for multimedia services, thus, they require more complex techniques, e.g., turbo codes in the HPAV and IEEE 1901 standards \cite{
IEEE1901}, while LDPCs in ITU-T G.9960 (G.hm). Besides the above mentioned techniques, some other coding schemes are currently under investigation by the research community. For instance, the permutation trellis codes which combine permutation and convolutional codes are particularly suited to combat burst of errors \cite{Ferreira}.

The modulation scheme is also important, especially because, as we saw in Section~\ref{sec:bbplcreg}, spectral masks have to be fulfilled as specified by the standards. Thus, it is important to realize flexible spectrum usage with the ability to create spectral notches and allow coexistence with other systems. The most commonly used modulation scheme is pulse-shaped OFDM (PS-OFDM), a multicarrier scheme similar to OFDM, but with the usage of a window which is better than the rectangular time-domain window adopted in OFDM. PS-OFDM is deployed in latest NB and BB PLC standards (see Section~\ref{sec:PLCstandardization}). It is also interesting to note that while BB system use coherent modulation, i.e., M-PSK and M-QAM, NB systems use also differential PSK. In particular, ITU-T G.9903 (G3-PLC) deploys the conventional time-differential phase modulation, while ITU-T G.9904 (PRIME) deploys frequency differential phase modulation where the information is encoded in the phase difference between adjacent 
OFDM sub-channels.

Recently, in view of an evolution for further improvements, more attention has been directed to the study of other types of filter bank modulation (FBM) that privilege the frequency confinement of the sub-channel pulses, e.g. filtered multitone (FMT) modulation \cite{Tonello_Pecile_Efficient}. FBM, such as FMT, offers several advantages over PS-OFDM, as the higher sub-channel frequency confinement and the higher notching selectivity, allowing a reduction in the number of sub-channels required to be deactivated to meet EMI constraints. In order to reduce the FBM implementation complexity, the use of a different architecture, where the linear convolutions are replaced with circular convolutions, was proposed in \cite{CBFMT_ISPLC}. In this case, the transmission takes place in blocks, similarly to OFDM, resulting in a scheme referred to as cyclic block FMT (CB-FMT). The circular convolution is also applied in the filter bank analysis at the receiver, offering an efficient frequency domain implementation \cite{
Tonello_2014}.

\begin{figure}[tb]
\centering
\includegraphics[width=\figurewidth]{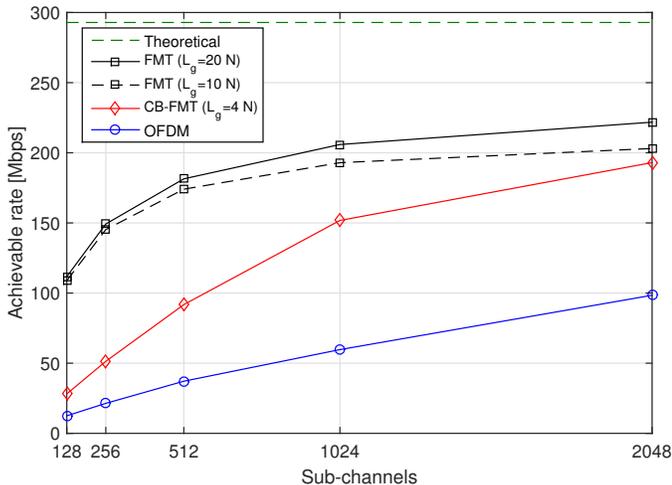}
\caption{Comparison between OFDM, FMT and CB-FMT in terms of maximum achievable rate as a function of the number of sub-channels used $K$.}
\label{fig:FMT_CBFMT_OFDM_cap}
\end{figure}
As an example, the achievable rate for a specific channel realization (which corresponds to the median channel, ranked in terms of capacity, selected from the database of measurements in \cite{TCOM1}) is reported in Fig.~\ref{fig:FMT_CBFMT_OFDM_cap} as a function of the number of used sub-channels. Different lengths $L_g$ of the prototype filter at the transmitter side are considered. The cyclic prefix (CP) length is chosen in order to offer the highest rate, under the further constraint of fulfilling the notching mask. The figure shows that the rate in CP-OFDM increases with the number of sub-channels. This happens since the overhead introduced by the CP is reduced and a better notching capability is obtained. However, the performance gap from the theoretical channel capacity is high. This gap can be significantly reduced by deploying FMT. In this example, FMT uses a long root-raised-cosine pulse with length 20 or 10 sub-channel symbols.  To reduce the FMT complexity, the figure shows that CB-FMT with a 
rectangular frequency domain pulse (which renders the system to be the dual of OFDM) provides better performance than PS-OFDM and, with 2048 sub-channels, it is not far away from FMT, despite the fact that its complexity is only 36 \% of the FMT one.

\subsection{PHY Layer Improvements}
\label{ssec:PHYimprov}
There are several areas that are currently being investigated for the improvement of the PHY layer. The main ones are: channel coding and signal processing for mitigating impulsive noise and interference \cite{Ferreira,Andreadou}; synchronization, channel estimation and equalization \cite{Cortes_07,lampe2015power}; transmission schemes that can allow coexistence at the PHY layer between different PLC systems \cite{GalliCOEX}, between high speed PLC systems together with sensor PLC networks \cite{TonelloIPLC} and between PLC systems and DSL systems \cite{Maes,Ali_14}; adaptation and resource allocation for maximum spectral efficiency, e.g., bit loading \cite[Ch. 6]{lampe2015power} or adaptation and allocation of the time/frequency resources in multicarrier systems \cite{TonelloCP,TonelloSLOT}; cooperation and relaying to extend coverage \cite{LampeREL,TonelloREL}; diversity combining techniques that mix PLC with wireless transmission \cite{Lai}.

In NB PLC for smart grid applications, research effort is spent to increase robustness and coverage, for instance, looking at increased spectrum usage, possibly beyond 148 kHz, or exploiting better coupling techniques that can resolve the limitations due to the low line impedance and high noise \cite{SGC_15}. In BB PLC, one question is how to go beyond the current high throughput offered to support very high speed multimedia services and home networking applications. One direction is to better exploit MIMO and another is to increase the bandwidth using an EMC friendly mechanism, for instance, as recently introduced by the HomePlug AV2 standard \cite{LL-yonge:2013}.

\section{Link and Higher Layers}\label{sec:mac} % David/Cristina

As we have seen, while the PLC medium has differences to the wireless
medium, it also has many similarities.  In particular it is a shared medium
with unpredictable channel behavior. Consequently, the link layer
and other layers immediately above face similar challenges to a
protocol designed for wireless. As we will see, this means that
the link layer and routing layer often have common elements with
their wireless counterparts. There are also differences: for example,
as mentioned above the PLC medium can be time-varying with the AC
mains cycle, and so transmissions or beaconing are aligned
with the mains cycle.

  \subsection{MAC Protocols for BB Applications}

The MAC protocols defined in the HomePlug \cite{HomeplugStd}, IEEE 1901 \cite{IEEE1901} and G.hn (G.9961) \cite{LL-ITU9961} standards define a contention-based (random access) as well as a contention-free (TDMA-like) procedure to access the channel. However, it is the random access procedure that usually is observed for Internet traffic \cite{cano2015icc}, and thus we will focus on this in the following. Although the three standards differ on how the channel access is realized, the random access procedure is equivalent. The general procedure adopted is a CSMA/CA technique, similar to the Distributed Coordination Function (DCF) defined in the IEEE 802.11 standard \cite{IEEE80211-IEEESTD1999}. Each time a node has a new packet to transmit, the backoff stage ($i \in [0,m-1]$) is initialized to $0$ and a random backoff counter (BC) is selected from $[0,W_{0}]$. The backoff countdown is frozen when activity is detected on the channel and restarted when the medium becomes idle again. The packet is actually 
transmitted when the backoff 
countdown expires. If an acknowledgment is received, the packet is considered successfully transmitted. Otherwise, the node starts the retransmission procedure: the backoff stage changes to $i=\min(i+1,m-1)$ and a new BC is selected from $[0,W_{i}]$, $W_{i}$ being the contention window of stage $i$. The similarity of the MAC to 802.11's DCF has lead to MAC modeling (e.g. \cite{chung2006performance,kriminger2011markov,cano2013PLCmodel,vlachou2014performance}) often in the style of Bianchi's 802.11 model \cite{bianchi2000performance}.
  
      \subsubsection{The Deferral Counter - Its Impact on Fairness}
  
In contrast to the DCF specification, in the HomePlug, IEEE 1901 MAC and G.hn (G.9961), a new counter, called the Deferral Counter (DC), is introduced. This counter is initialized to $M_i$ at each backoff stage and decreased by one after overhearing a data packet or a collision. If a new packet or a collision is overheard and the value of the DC is equal to zero, the node acts as if a collision has happened: the backoff stage is increased if it has not yet reached its maximum value and a new backoff is selected from $[0,W_{i}]$. The goal of the DC is to avoid collisions when high contention is inferred by decreasing the aggressiveness of transmission attempts.

The use of the deferral counter does reduce the collision probability when there is high contention. However, as shown in \cite{cano2013pimrc}, this modification to the DCF does not always provide better performance, especially considering heterogeneous and exposed terminal scenarios. More importantly, it has been shown in \cite{vlachou2013fairness} that it has an impact on short-term fairness as some stations may substantially reduce their transmission probability by overhearing consecutive neighboring transmissions at a given time interval. The trade-off between collision probability and fairness has been studied in \cite{vlachou2014new}.
 
    \subsubsection{Strict Prioritization - Benefits and Drawbacks}
    
\begin{figure}[!tb]
\centering
\includegraphics[width=3.4in]{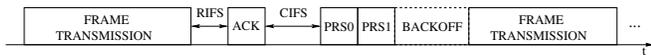}
\caption{Allocation of priority resolution slots (refer to \cite{HomeplugStd} and \cite{IEEE1901}).}
\label{fig:prs_scheme}
\end{figure}

To provide channel access differentiation, HomePlug \cite{HomeplugStd}, IEEE 1901 \cite{IEEE1901} and G.hn (G.9961) \cite{LL-ITU9961} define four Access Categories (CAs), ranging from CA0--3. CA3 and CA2 share $W_i$ and $M_i$ values, as do CA1 and CA0. Two Priority Resolution Slots (called PRS0 and PRS1) are allocated at the end of successful frame exchanges, see Fig. \ref{fig:prs_scheme}. These slots allow nodes to announce the priority of packets pending transmission. The highest priority (CA3) is signaled by transmitting a symbol in both PRS0 and PRS1; the CA2 category is signaled in PRS0 only; CA1 signals in PRS1, if PRS0 was empty; and the lowest access category (CA0) does not signal at all. Following this approach, stations know if there is a station with a frame that belongs to a higher CA. In such a case, they do not contend for the channel, allowing high-priority frames to be released.

This resolution scheme aims to provide strict access differentiation, i.e. using the priority resolution mechanism, packets with higher priority are always transmitted before lower-priority ones. However, the priority resolution scheme in HomePlug and IEEE 1901 is only invoked after successful frame exchanges. These standards suggest that PRS are not present after: \emph{i)} a collision, \emph{ii)} frame transmissions resulting in erroneous receptions and \emph{iii)} the detection of an empty channel for longer than an Extended InterFrame Space (EIFS) period. Thus, in lightly loaded conditions and after collisions or channel errors, the priority resolution scheme is not employed and channel access differentiation only occurs through the different parameters of the access categories.  Thus, we expect strict prioritization if we have a single station in a high CA, but less strict prioritization if multiple stations are in the highest CA because of collisions.

\begin{figure}[!tb]
\centering
\includegraphics[width=2.9in]{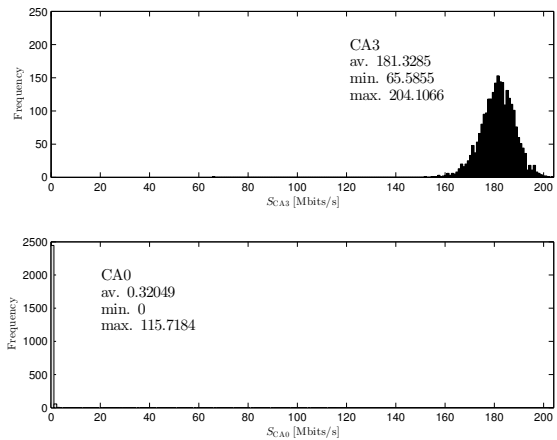}
\caption{Histrograms of throughput for 2 stations configured at CA3 and CA0. Experimental results extracted from \cite{cano2015icc}.}
\label{fig:results_1ca3_1ca0}
\end{figure}

Channel differentiation in PLC networks has been evaluated in \cite{chowdhery2009polite,zarikoff2011construction,meftah2011ns,cano2014isplc} and \cite{cano2015icc}. Although the priority resolution mechanism is able to provide strict protection to high-priority traffic, a series of issues has been identified. First, one immediate effect of this strong protection to high-priority frames is the starvation faced by lower-priority traffic \cite{cano2015icc}. To illustrate the importance of this effect we show in Fig. \ref{fig:results_1ca3_1ca0} histograms of throughput from experimental results in a real testbed extracted from \cite{cano2015icc}. Note that in presence of a higher access category, the station configured at CA0 is not effectively able to transmit. This effect is caused by the inability to transmit any lower-priority frame when stations with higher-priority frames are saturated, i.e. they always have a packet pending to transmit in their queues. Moreover, as shown in \cite{cano2014isplc}, the behavior 
of the network is extremely hard to predict when we vary the number of stations contending for the channel or when the traffic load changes. Additionally, given that control messages for tone map update are sent at CA2, a possible oscillatory behavior in throughput has been identified \cite{
cano2015icc}. This is because it is impossible to release these control messages in presence of a saturated CA3 source, which prevents a given station transmitting as its tone map is considered stale.

    \subsubsection{Aggregation and Buffer Management - Efficiency vs. Variability}
    
The standards HomePlug \cite{HomeplugStd}, IEEE 1901 \cite{IEEE1901} and G.hn (G.9961) \cite{LL-ITU9961} provide a high degree of aggregation capabilities. Experimental studies on commercial IEEE 1901 devices show that although the efficiency can be improved by aggregating more data in a single transmission when channel conditions are favorable, it may result in a high degree of variability on performance \cite{cano2015icc}. A challenging aspect of studying this is that both aggregation techniques and buffer management decisions are vendor-specific.
    
  \subsection{MAC Protocols for NB Applications}

The MAC protocols for NB applications share many similarities but they also have several differences that are important to consider \cite{LL-galli:2015}. We give an overview of their common characteristics and differences next. 

The standard PRIME defines a Contention Free access Period (CFP) in which devices transmit using TDMA, as well as a Contention Access Period (CAP) where channel accesses are randomized. During CAP, stations wait a random backoff before attempting transmission. The random backoff depends on the priority of the frame and on the number of experienced channel attempts. After the backoff expires, stations carrier sense the medium a number of times if it is found busy. The number of times the carrier sensing procedure is carried out also depends on the priority of the frame to be transmitted. 

Contrary to PRIME, the MAC protocol defined in IEEE 1901.2 is based on the IEEE 802.15.4 standard for wireless sensor networks \cite{IEEE802154-IEEESTD2003}. The same applies to G3-PLC, except for the fact that G3-PLC does not define a CFP period. In G3-PLC and in the CAP period of IEEE 1901.2, stations perform a random backoff before attempting transmission, similarly to IEEE 802.15.4. G3-PLC and IEEE 1901.2 extend the contention procedure of IEEE 802.15.4 to account for fairness (stations with a high number of busy channel detections increase the aggressiveness of their transmission attempts) and different priorities (by defining different contention periods for different access categories). A common feature of the random access procedure in G3-PLC, IEEE 1901.2 and PRIME is that the assessment of the channel status only occurs when the backoff expires. 

Finally, the NB MAC procedure defined in G.hnem is more like the CSMA/CA approach defined in the IEEE 802.11 standard \cite{IEEE80211-IEEESTD1999} for wireless local area networks. In this case, no CFP is defined, and thus stations access the channel following a contention-based approach. If a transmission is detected on the channel, stations defer their attempt until the next contention period, which takes place once the current ongoing transmission is completed. G.hnem also accounts for different prioritization levels. However, in contrast to G3-PLC and IEEE 1901.2, the differentiation is not so strict as different access categories have contention windows that partially overlap.

Although the PLC research community can rely on the extensive work on IEEE 802.15.4 and IEEE 802.11 to predict the performance of the network, these standards include some modifications which are, as far as we know, unexplored at present. These are: \emph{i)} the modifications in G3-PLC and IEEE 1901.2 of the backoff procedure to provide fairness, \emph{ii)} the strict prioritization mechanisms defined in G3-PLC and IEEE 1901.2, and \emph{iii)} the prioritization mechanism defined in G.hnem. The impact of these extensions on performance is not straightforward and further analysis is needed in order to understand the behavior of the network.

  \subsection{Routing Issues}

G.hn and IEEE 1901 support link layer multihop operation, where nodes that
are not in direct communication can have frames received and
retransmitted via intermediate nodes, and the protocol
can take advantage of link quality information provided by the lower
layers. There is also a possibility to take advantage of other
aspects of PLC networks for routing. For example, previously we discussed
that the topology often has a tree-like structure, which might be
exploited by a routing system \cite{zhang2013novel}. Likewise, for
many PLC devices, it is likely that they are attached at a physically
fixed location, and so geographic routing may be practical
\cite{biagi2010location,biagi2012neighborhood}.

Indeed the NB PLC standards are divided on whether routing should be
carried out at the link layer or above
\cite{LL-Bookcollection:PLC-Galli-2014Chap11,galli2015next}.  G.9903
and G.9904 include link-layer routing, where all nodes appear to
be connected, even if relaying is taking place.  In contrast, IEEE
1901.2 and G.9902 allow routing at the link layer or above, where
in the latter case a higher layer protocol must handle forwarding
between nodes not in direct communication.

Of course, rather than receive and retransmit, it is possible to
have nodes relay in real-time, as in cooperative transmission in
wireless.  Cooperative relaying has been considered for PLC (e.g.
\cite{kuhn2006power,Lampe_12,TonelloREL,LampeREL,cheng2013relay}). While
diversity gains are often lower than in wireless, power gains through
multihop transmission are still practical, which can be useful for
improving range.

  \subsection{Integration with the Networking Ecosystem}

Some integrations of PLC into the broader network have been successful.
Consumer modems, or integrated PLC-WiFi devices for extending the
reach of networks are available off-the-shelf
\cite{tinnakornsrisuphap2014coverage}. The IEEE 1905.1 standard provides
a convergence layer to facilitate the use of WiFi, PLC, Ethernet and
MoCA within the home \cite{IEEE1905.1,LL-Bookcollection:PLC-Lampe-2015Chap15}. A generic extension mechanism has also been recently standardized in IEEE 1905.1a \cite{IEEE1905.1a}.
Other uses of PLC have been
proposed, for example a mix of WiMAX and PLC has been proposed for
collecting data in a hospital \cite{wang2010hybrid}. Visible light
communication is another promising technology that is complementary
to WiFi and PLC \cite{ma2013integration}.

Another question is what protocol should be run over
PLC. Broadband PLC is often used like Ethernet, and so can be used
in much the same way as any LAN, running IPv4, IPv6 or other protocols.
However, in NB PLC, sometimes resources are at more of a premium.
For example, in G.9903 6LoWPAN is used to carry IPv6 frames on the
PLC network. 6LoWPAN provides a number of functions that might
be optimized specifically for PLC (e.g. routing \cite{razazian2013enhanced}).

Of course, when integrated with the network ecosystem, PLC needs
to be managed. A number of PLC vendors provide tools, including
open source tools such as faifa\footnote{Available at \url{https://github.com/ffainelli/faifa}}
and open-plc-utils\footnote{Available at \url{https://github.com/qca/open-plc-utils}}. Efforts have
also been made to provide an SNMP interface to PLC devices
\cite{park2008definition,park2010development,de2010management}.
Within hybrid networks, IEEE 1905.1's convergence layer also provides
abstractions to help with management, including features for
estabilishing the topology and link metrics \cite{IEEE1905.1,LL-Bookcollection:PLC-Lampe-2015Chap15,IEEE1905.1a}.

  \subsection{Challenges and Future Directions}
  
Compared to research on IEEE 802.11 and IEEE 802.15.4 networks, and also compared to the advances on the physical layer of PLC networks, the MAC protocols for PLC are unexplored. Work has begun to fill this gap, however there are still many aspects that remain unclear and several issues that need to be studied in order to ensure the successful penetration of the technology. 

In particular, extensions to the analytical models of HomePlug and IEEE 1901 \cite{vlachou2014performance,vlachou2014new,cano2013PLCmodel} in order to consider aggregation and buffer management techniques are needed in order to fully understand the protocol behavior and the impact on performance. Similarly, amendments to the standards related to the deferral counter and the strict priority resolution scheme may also be desirable. Also, as previously stated, the impact on performance of the extensions to the IEEE 802.15.4 and IEEE 802.11 baselines considered in NB PLC standards regarding fairness and prioritization remain relatively unstudied.

A combined understanding of PHY effects and MAC layer issues can raise interesting issues.  An example of this is the challenge of building a stream protocol for smart grid on top of stop-and-wait MAC protocols common in PLC \cite{bumillerpowerline}, where MAC delays can indirectly result from using long OFDM symbols to mitigate burst interference. Though IEEE 1905.1 provides basic mechanisms, such as link metrics, for addressing the use of hybrid PHY layers, the optimal use of its routing and multipath forwarding features are still open questions.

%\section{Industrial Perspectives}\label{sec:industry} % Anand?
%
%  \subsection{Industrial Roadmap}
%  \subsection{Availability of chips and devices}
  
\section{Final Remarks}\label{sec:remarks}

PLC networks have become a fruitful technology which can provide a means of communication for a wide range of applications. There have been numerous regulatory and standardization efforts over recent decades and a programme of work by the research community has addressed different challenges, making great advances on the use of a channel not initially designed for data communication. 

In this article we have reviewed germane contributions and stated the main results in the literature, for both NB and BB systems. We have considered standardization, channel characterization and modeling, as well as physical and higher layer techniques defined in the different PLC standards.  

We have also highlighted areas of further study. Regarding the physical layer, we have pointed out future research directions that include channel coding and signal processing, mechanisms to ensure coexistence among different PLC systems and among PLC and other communication technologies, resource allocation in multicarrier systems, techniques to extend coverage based on cooperation and relaying, and combining the use of PLC with wireless transmission using diversity-combining techniques. On the higher layers, we have emphasized the need for further studies on the differences of the protocols defined for PLC networks compared to their analogs for wireless or sensor networks, mechanisms to resolve effects due to strict priority resolution, combined behavior of the medium access control and the physical layer dynamics and the integration with the networking ecosystem. We believe research outcomes in these areas will increase the penetration of PLC in the years to come.

%\section*{Acknowledgments}

\ifCLASSOPTIONcaptionsoff
  \newpage
\fi

\bibliographystyle{IEEEtran}
\bibliography{library}

\newpage

\begin{wrapfigure}{l}{30mm}
    \includegraphics[width=30mm]{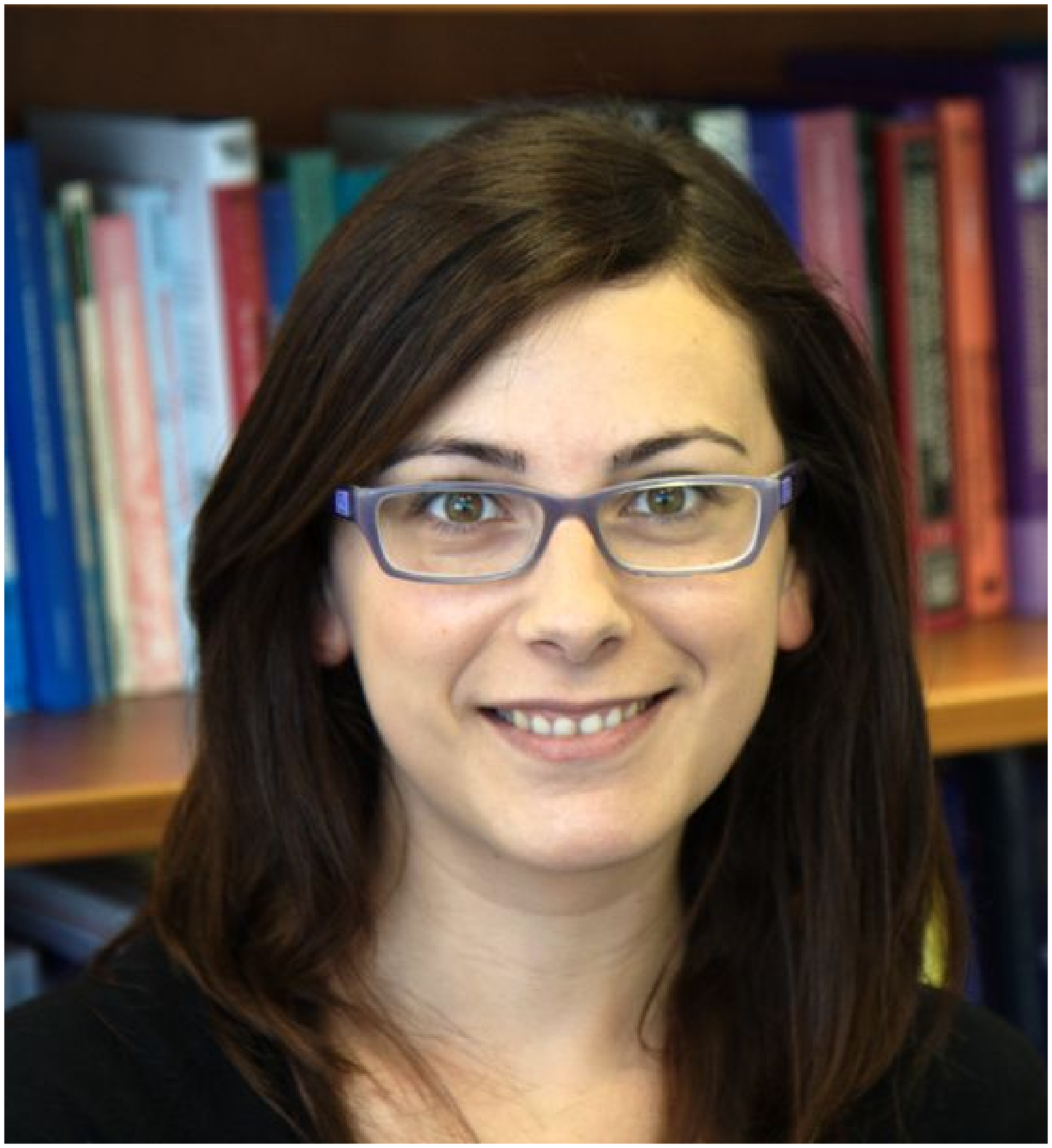}
\end{wrapfigure}
\textbf{Cristina Cano} obtained the Telecommunications Engineering Degree at the Universitat Politecnica de Catalunya (UPC) in February 2006. Then, she received the M.Sc. (2007) and Ph.D. (2011) on Information, Communication and Audiovisual Media Technologies from the Universitat Pompeu Fabra (UPF). From July 2012 to December 2014, she was working as a research fellow at the Hamilton Institute in the National University of Ireland, Maynooth (NUIM). Part of this research group is now based in Trinity College Dublin (School of Computer Science and Statistics), where she was working from January 2015 to January 2016 as a senior research fellow. Now she is with the FUN team at Inria-Lille. Her research interests are related to coexistence of wireless heterogeneous networks, distributed scheduling, performance evaluation and stability analysis. \\

\begin{wrapfigure}{l}{30mm}
    \includegraphics[width=30mm]{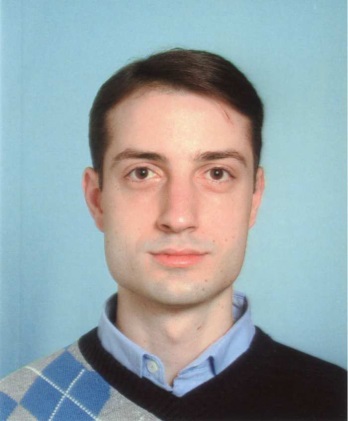}
\end{wrapfigure}
\textbf{Alberto Pittolo} received the Laurea degree in electrical engineering (2009) and the Laurea Specialistica degree (2012) in electrical and telecommunications engineering, with honors, from the University of Udine, Udine, Italy. He is currently completing the Ph.D. degree in industrial and information engineering at the University of Udine. His research interests and activities are channel modeling, physical layer security and resource allocation algorithms for both wireless and power line communications.
\\

\begin{wrapfigure}{l}{30mm}
    \includegraphics[width=30mm]{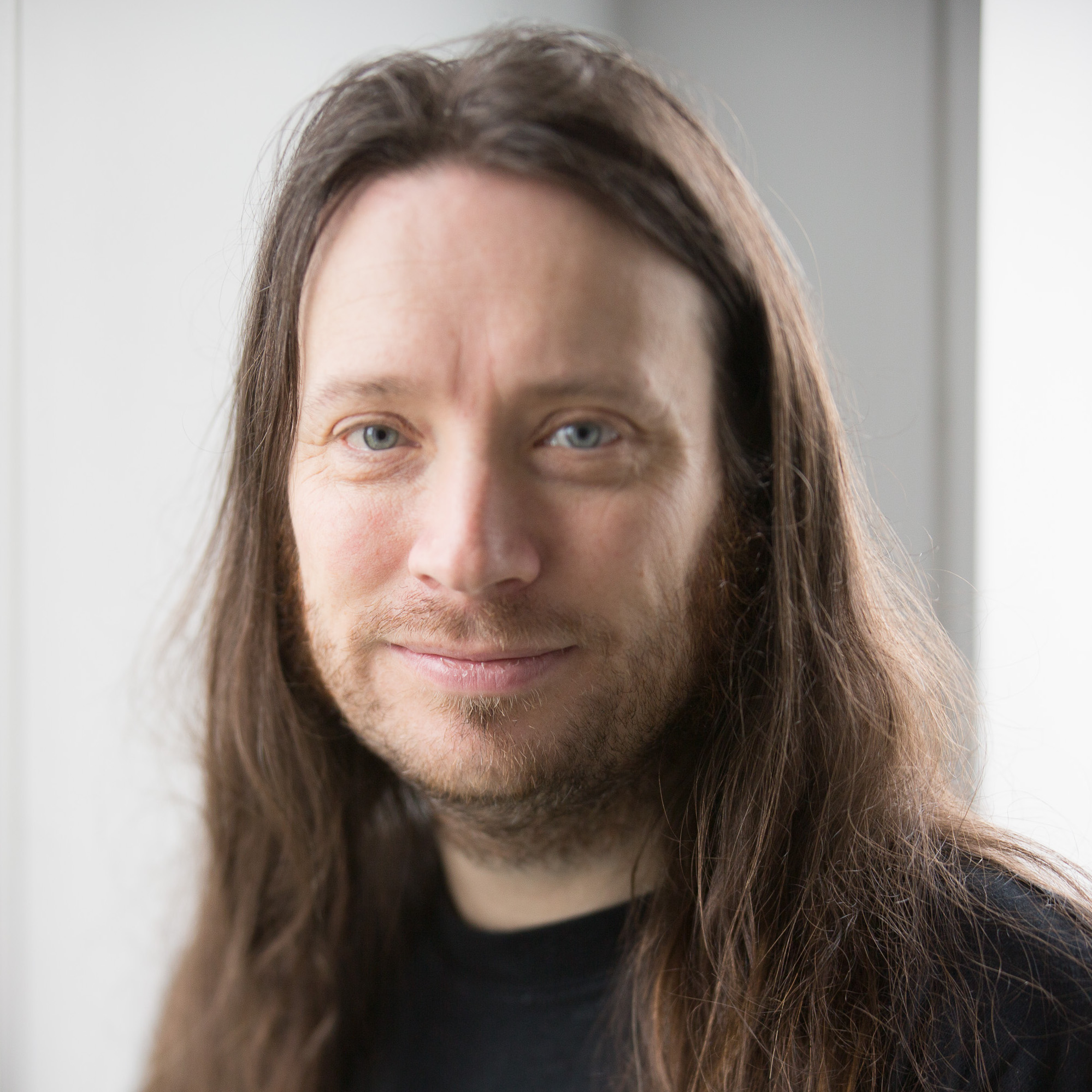}
\end{wrapfigure}
\textbf{David Malone} received B.A.(mod), M.Sc. and Ph.D. degrees in mathematics         
from Trinity College Dublin. During his time as a postgraduate, he              
became a member of the FreeBSD development team.  He is currently a                
member of Maynooth University's Hamilton Institute and Department of               
Mathematics \& Statistics. His interests include mathematical                   
modeling and measurement of WiFi, PLC and password use. He also                 
works on IPv6 and systems administration, and is the is a co-author             
of O'Reilly's ``IPv6 Network Administration''.\\

\newpage

\begin{wrapfigure}{l}{30mm}
    \includegraphics[width=30mm]{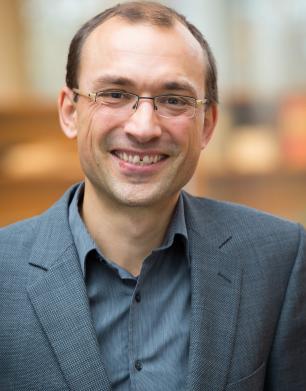}
\end{wrapfigure}
\textbf{Lutz Lampe} (M'02–SM'08) received the Dipl.-Ing. and Dr.-Ing. degrees in electrical engineering from the University of Erlangen, Erlangen, Germany, in 1998 and 2002, respectively. Since 2003, he has been with the Department of Electrical and Computer Engineering, The University of British Columbia, Vancouver, BC, Canada, where he is a Full Professor. His research interests are broadly in theory and application of wireless, optical wireless, and power line communications. Dr. Lampe was the General (Co-)Chair for the 2005 International Conference on Power Line Communications (ISPLC), the 2009 IEEE International Conference on Ultra-Wideband (ICUWB) and the 2013 IEEE International Conference on Smart Grid Communications (SmartGridComm). He is currently an Associate Editor of the IEEE WIRELESS COMMUNICATIONS LETTERS and the IEEE COMMUNICATIONS SURVEYS AND TUTORIALS and has served as an Associate Editor and a Guest Editor of several IEEE transactions and journals. He was a (co-)recipient of a number of best paper awards, including awards at the 2006 IEEE ICUWB, the 2010 IEEE International Communications Conference (ICC), and the 2011 IEEE ISPLC.\\

\begin{wrapfigure}{l}{30mm}
    \includegraphics[width=30mm]{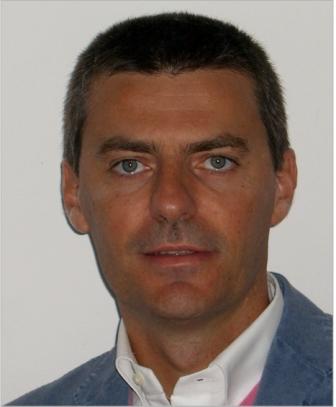}
\end{wrapfigure}
\textbf{Andrea M. Tonello} (M00-SM12) received the Laurea degree (summa cum laude) and the PhD in electronics and telecommunications from the University of Padova, Italy, in 1996 and in 2003, respectively. From 1997 to 2002, he was with Bell Labs- Lucent Technologies, Whippany, NJ, USA, first as a Member of the Technical Staff. Then, he was promoted to Technical Manager and appointed to Managing Director of the Bell Labs Italy division. In 2003, he joined the University of Udine, Italy, where he became Aggregate Professor in 2005 and Associate Professor in 2014. He also founded WiTiKee, a spin-off company working on telecommunications  for smart grids. Currently, he is Professor and Chair of the Embedded Communication Systems Group at the University of Klagenfurt, Austria. Dr. Tonello received several awards, including five best paper awards, the Bell Labs Recognition of Excellence Award (1999), the Distinguished Visiting Fellowship from the Royal Academy of Engineering, U.K. (2010), the IEEE VTS Distinguished Lecturer Award (2011-2015), the Italian Full Professor Habilitation (2013). He is the Chair of the IEEE Communications Society Technical Committee on Power Line Communications. He serves as Associate Editor of IEEE Trans. on Communications and IEEE Access. He was the General Chair of IEEE ISPLC 2011 and IEEE SmartGridComm 2014.\\

\begin{wrapfigure}{l}{30mm}
    \includegraphics[width=30mm]{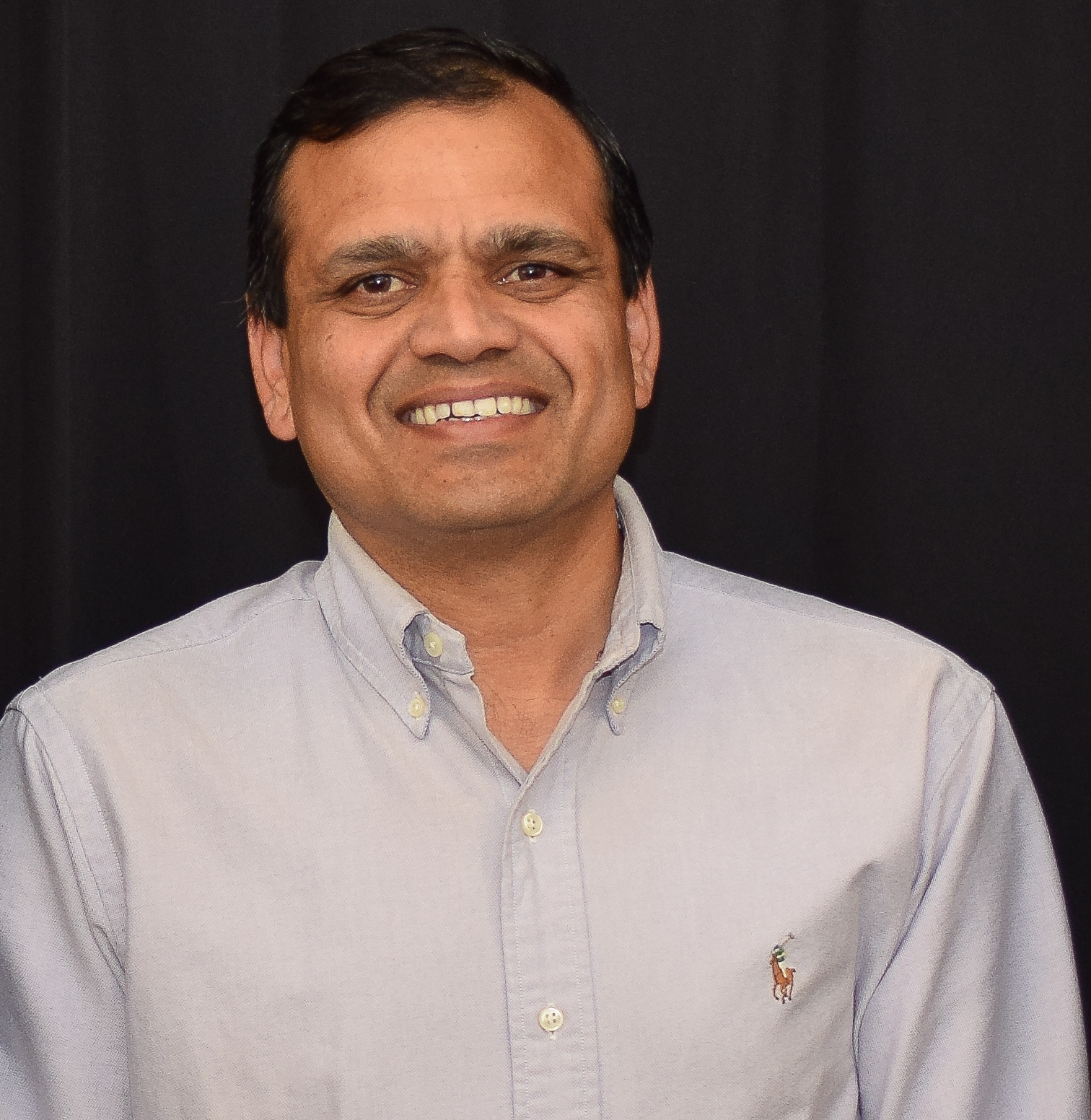}
\end{wrapfigure}
\textbf{Anand G. Dabak} received the B.Tech. degree in electrical engineering from the Indian Institute of Technology, Bombay, India in 1987 and the M.Sc. and Ph.D. degrees in electrical engineering, from Rice University, Houston, TX, in 1989 and 1992, respectively in statistical signal processing. After receiving the Ph.D. degree, he worked for Viasat Inc., Carlsbad, California, before joining Texas Instruments Incorporated, Dallas, TX, in 1995. He has since then worked on the system and algorithm issues related to communication systems including wireless and powerline communications (PLC). He was involved in wireless standardization and chipset solution development activity for GSM, 3G, LTE, Bluetooth systems and PLC standardization in IEEE 1901.2 and solutions on TI MCU platforms. He is currently IEEE Fellow and TI fellow and continues to work Ultrasound based water and gas flowmetering and other signal processing applications that can be implemented on TI platforms. He has published more than 30 journal and conference papers and has more than 100 issued patents.\\

\end{document}